\documentclass[pra,floatfix,amsmath,amsfonts, twocolumn]{revtex4}
\usepackage{graphicx}
\usepackage[utf8]{inputenc}
\usepackage{dcolumn}
\usepackage{bm}
\usepackage{amsmath}
\usepackage{amssymb}
\usepackage{wasysym}
\usepackage{color}

\begin{document}

\title{Photonic Nambu-Goldstone bosons}
\author{Miguel-\'Angel Garc\'\i a-March,$^1$ Angel Paredes,$^2$,
Mario Zacar\'es,$^3$  Humberto 
Michinel$^2$ and Albert Ferrando,$^4$ }
\affiliation{
$^1$ ICFO – Institut de Ci\`encies Fot\`oniques, The Barcelona Institute of Science and Technology, 08860 Castelldefels (Barcelona), Spain.\\
$^2$Departamento de F\'\i sica Aplicada, Universidade de Vigo, As Lagoas s/n, Ourense, ES-32004 Spain.\\
$^3$ Facultad de Veterinaria y Ciencias Experimentales, Universidad Cat\'olica de Valencia, `San Vicente Martir", (Val\`encia), Spain.\\
$^4$ Departament d'\`Optica i Optometria i Ci\`encies de la Visi\'o. Interdisciplinary Modeling Group, {\it InterTech}. Universitat de
Val\`encia, Burjassot (Val\`encia), Spain.
}

\begin{abstract}

We study numerically the spatial dynamics of light in periodic square lattices in the presence of a Kerr term,
emphasizing the peculiarities stemming from the nonlinearity. 
We find that, under rather general circumstances, the phase pattern of the stable ground state
depends on the character of the nonlinearity:  the phase is spatially uniform if it is defocusing whereas 
in the focusing case, it presents a chess board pattern, with a difference of $\pi$ between neighboring
sites. We show that the lowest lying perturbative excitations can be described as perturbations
of the phase and that finite-sized structures can act as tunable  
metawaveguides for them. 
The tuning is made by varying the intensity of the light that, because of the nonlinearity,
affects the dynamics of the phase fluctuations.
We interpret  the results using methods of condensed matter physics, based on an effective
description of the optical system. This interpretation sheds new light on the phenomena, facilitating
the understanding of individual systems and leading to a framework for relating different problems
with the same symmetry.
In this context, we show that the perturbative excitations of the phase are Nambu-Goldstone bosons
of a spontaneously broken $U(1)$ symmetry.

\end{abstract}


\maketitle


\section{Introduction}
\label{sec:intro}

Nonlinearities have played a key role
in discovering an amazing wealth of phenomena in laser beam propagation,
in the manipulation of light and in establishing connections between optics and
other areas of physics. Many analogies with condensed matter
physics have been presented in the literature, see e.g.
 \cite{morandotti1999experimental,freedman2006wave,schwartz2007transport}.
Optical versions have appeared
for different kinds of spatially ordered  lattices as, e.g., honeycomb graphene-like configurations \cite{peleg2007conical}
or Lieb lattices \cite{leykam2012pseudospin}. 
Remarkable
concepts like the topological protection of transport properties have been translated from 
the condensed matter community to photonic systems
 \cite{raghu2008analogs,lu2014topological,plotnik2015topological}.
 In most situations,
the underlying photonic structure is a linear or nonlinear refractive index which satisfies
certain periodicity conditions along the plane transverse to beam propagation. 
It naturally
induces an ordering in the distribution of light which can be (partially) identified with
the wavefunction of electrons within crystals. 
It has been recently shown that the spatial
ordering can also arise spontaneously for appropriately chosen nonlocal nonlinearities
\cite{maucher2016self}.

In this context, 
the present work deals with  pattern formation and 
light propagation in nonlinear lattices \cite{kartashov2009soliton,kartashov2011solitons}.
We consider  square lattices, a symmetry that has been studied in relation to guiding of 
light \cite{bouk2004dispersion}, formation of discrete solitons \cite{fleischer2003observation}, 
supercontinuum generation
\cite{begum2011supercontinuum}, lasers \cite{liang2011three} and
metamaterials \cite{huang2011dirac},
just to mention a few examples.
Moreover, 
it is well known that the nonlinear Schr\"odinger equation on which our work is grounded
is also the mean-field description of  Bose-Einstein condensates
\cite{dalfovo1999theory}. Thus, our results can also find application in the semiclassical
modeling of cold atoms in optical lattices \cite{lewenstein2007ultracold}.

\begin{figure}[h!]
\begin{center}
\includegraphics[width=\columnwidth]{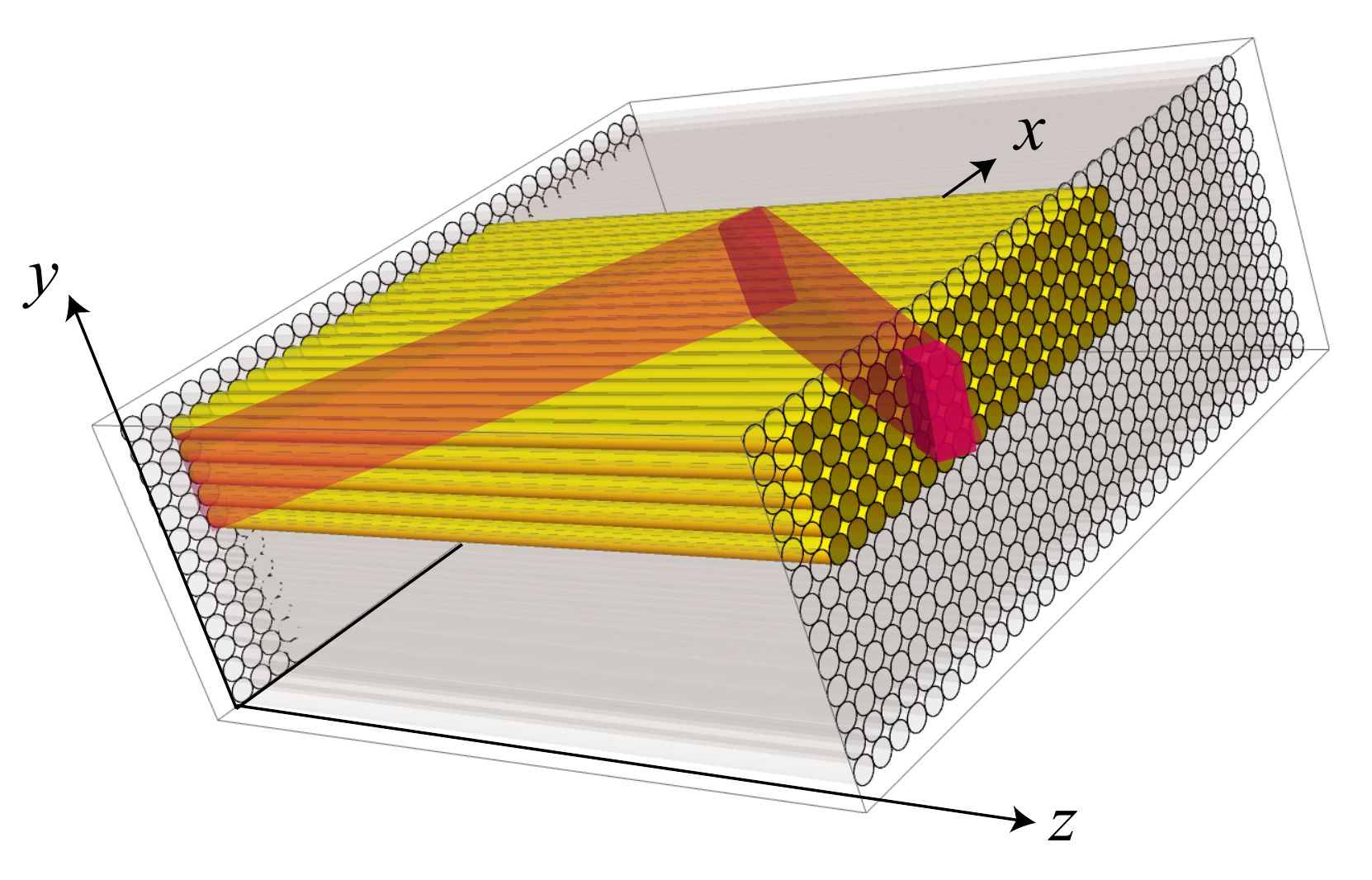}
\end{center}
\caption{
The black circles represent a square photonic lattice, within which 
a finite size nonlinear solution for the optical beam propagates (yellow structure). With adequate features,
 it acts as an effective metawaveguide for phase 
 excitations of the optical field (red path). These phase 
 perturbations behave as photonic Nambu-Goldstone bosons. }
\label{fig0}
\end{figure}

Concretely, we study a case in which the linear refractive index varies harmonically and
the nonlinear refractive index is taken as a constant Kerr term. The sign of the nonlinear
term is decisive for the character of the ground eigenstate, since the phase pattern
depends on it. In the defocusing case, the phase is spatially constant. In the focusing
case, it has opposite sign in neighboring sites, forming a chess board pattern.
Even if the computations are performed numerically for particular examples, we argue that
this conclusion is rather general and depends on the symmetry and the interplay between
the nonlinearity and the lattice. After discussing
the diffractionless modes and their stability, we study perturbations of the ground state.
Of particular interest are those sparked by altering the phase of some sites of the lattice.
The subsequent evolution can be (approximately) interpreted as a wave that transports the
introduced phase difference while leaving the intensity pattern (almost) unchanged.
We  show how finite size
propagating nonlinear beams inside a photonic lattice can act as effective ``metawaveguides" for these 
 phase perturbations, as depicted in Fig.~\ref{fig0}. Remarkably, the propagation properties of the phase wave 
depend on the nature of the underlying nonlinear beam, which in this way behaves as an optical  
``metamaterial" whose susceptibility can be  tuned
by varying the power of the beam.

Apart from the discussion in terms of the usual formalism in nonlinear optics,
a major goal of this work is to introduce methods that are customarily used in the solid state
literature \cite{ashcroft1978solid} but that have not been fully exploited in the  present
framework.
We devise a mean field formulation, valid for
excitations of long wavelength compared to the lattice spacing.
The procedure is, in spirit, analogous to renormalization in 
condensed matter or particle physics. 
The resulting effective description for an order parameter allows us, using Landau theory,
to discuss the stability of the nonlinear modes and their lowest lying excitations.
The aforementioned perturbations of the phase turn out to be a photonic example
of a Nambu-Goldstone boson, related to the spontaneous symmetry breaking
of the $U(1)$ symmetry of phase rotations.
We emphasize that the introduced formalism can shed new light on results directly obtained from the 
complete model and is instrumental in formulating new predictions.

The paper is organized as follows: in Section \ref{sec:model} we introduce the mathematical
model and comment on its relation to nonlinear propagation in photonic lattices.
Section \ref{sec:eigenstates} provides a numerical discussion of the eigenstates and their
stability. In Section \ref{sec:effective}, we construct the effective approximate description,
using a formalism borrowed from the condensed matter literature.
In Section \ref{sec:goldstone}, we analyze the perturbative excitations 
and demonstrate their Nambu-Goldstone nature.
In Section \ref{sec:finite}, we study finite lattices and
 introduce
the concept of metawaveguide. 
 Finally, in Section \ref{sec:discussion} we conclude
and outline possible future directions.
In Appendix A, we introduce in a  detailed 
and pedagogical way the concepts of Bloch and Wannier functions,
for readers that might not be familiar with the formalism.
Appendix B gives a precise definition of what we define as the phase at each lattice cell.

\section{The model}
\label{sec:model}

In order to model paraxial nonlinear propagation of monochromatic light,
we consider the dimensionless nonlinear Schr\"odinger equation in the following form
\begin{equation}
i \frac{\partial \psi}{\partial z} = - \nabla^2 \psi + V(x,y) \psi - g |\psi|^2 \psi,
\label{eq1}
\end{equation}
where $g=+1$ or $g=-1$, depending on the focusing or defocusing character of the
Kerr nonlinearity and $\nabla^2$ is the two-dimensional Laplacian. 
The potential of the linear term is assumed to be periodic, defining cells of unit size
which form a square lattice,
\begin{equation}
V(x,y)=V(x+1,y)=V(x,y+1).
\label{eqVperiodic}
\end{equation}
For definiteness, computations will be performed using
\begin{equation}
V= V_0 \cos^2 (\pi x) \cos^2(\pi y).
\label{Vexample}
\end{equation}
As a convenient mathematical trick, customarily used in condensed matter physics, 
we also introduce periodic boundary conditions for the wavefunction
\begin{equation}
\psi(x,y)=\psi(x+N,y) = \psi(x,y+N),
\label{psiperiodic}
\end{equation}
for some integer $N\gg1$, such that there are $N^2$ unit cells in the lattice.
Obviously, the condition of Eq. (\ref{psiperiodic}) is unphysical. We will relax it in
Section \ref{sec:finite}, where finite structures are considered. 
Most of the features found using Eq. (\ref{psiperiodic}) apply to realistic setups.

This dimensionless formalism  is connected to
 photonic propagation in a periodic medium as follows.
The standard paraxial wave equation for a laser beam reads
\begin{equation}
-2ik_0 n_0\frac{\partial A}{\partial \tilde z}= \tilde \nabla^2 A + 2 \Delta n k_0^2 n_0 A,
\label{parax1}
\end{equation}
where 
$k_0=\omega/c=2\pi/\lambda$ is the wavenumber in vacuum and
 $A$ is the wave envelope, slowly varying at the scale of the wavelength.
Dimensionful spatial coordinates are denoted as $\tilde x$, $\tilde y$, $\tilde z$
and the laplacian is $\tilde \nabla^2 \equiv \partial^2_{\tilde x}+ \partial^2_{\tilde y}$.
The refractive index is $n=n_0 + \Delta n$ where $n_0$ is a constant and for
$\Delta n \ll n_0$ we consider the sum of a nonlinear component 
and a linear part associated to the inhomogeneity of the optical material
$\Delta n= \frac{n_2n_0}{2\eta_0}|A|^2 + \Delta n_{lin} (\tilde x,\tilde y) $, 
where $n_2$ is the nonlinear refractive index and
 $\eta_0 = \sqrt{\mu_0/\epsilon_0}$.
We now assume that $\Delta n_{lin} (\tilde x,\tilde y)$ is periodic in space, 
$\Delta n_{lin} (\tilde x,\tilde y)=\Delta n_{lin} (\tilde x+a,\tilde y)=\Delta n_{lin} (\tilde x,\tilde y+a)$
where  $ a$ is the  lattice spacing. 
We can rescale Eq. (\ref{parax1}) in order to bring it to dimensionless form, Eq. (\ref{eq1}),
by taking
\begin{eqnarray}
&&(\tilde x, \tilde y) = a (x,y)\ , \quad
A = \frac{1}{n_0 k_0 a} \sqrt\frac{\eta_0}{|n_2|} \psi, 
 \nonumber\\
 &&\tilde z= 2 k_0 n_0 a^2 z\ , \quad\ \
\Delta n_{lin} = - \frac{1}{2 n_0 k_0^2 a^2} V.
\end{eqnarray}
In particular, notice that $V$ scales with $a^2$ and therefore the value of $V_0$ 
can be tuned by adjusting the lattice spacing. For instance, imagine a sample value of
$n_0=1.5$, and a maximum deviation of the linear refractive index
 $|\Delta n_{lin}| = 0.1$. Then, the value  $V_0=800$ 
 that will be used in the illustratory examples of the following sections 
  corresponds to
$a \approx 8 \lambda$.
The relation between the dimensionless power and the physical one is
\begin{equation}
\tilde P= \frac{\lambda^2}{8 \pi^2 n_0|n_2|}P.
\end{equation}
As it is somewhat customary \cite{kivshar1998dark,stegeman1999optical,paredes2014coherent}, 
we will abuse language and identify the propagation distance 
$\tilde z$ with a ``time'' parameterizing the evolution. From this point of view, we
refer to the velocity
of a perturbation $\tilde v$ (which is an angle in physical terms)
as the displacement in $\tilde x$, $\tilde y$ divided by $\tilde z$. The relation to the
dimensionless velocity is 
\begin{equation}
\tilde v = \frac{\lambda}{4\pi n_0 a} v.
\end{equation}

The particular potential of Eq. (\ref{Vexample}) was chosen because of its simplicity and because it can 
be generated by illuminating a photorefractive material with a standard interference 
pattern~\cite{efremidis2002discrete}.
Equation~(\ref{eq1}) and therefore most of 
 our discussion can be applied to the dynamics of Bose-Einstein condensed cold atoms.
In that context, Eq.~(\ref{Vexample}) is a typical profile for two-dimensional 
optical lattices~\cite{morsch2006dynamics} although it is well-known that other potentials are
also possible~\cite{santos2004atomic}.  

The discussion and general conclusions of this work depend 
decisively on Eq.~(\ref{eqVperiodic}) but not
on the concrete expression Eq.~(\ref{Vexample}). 
The nonlinear refractive index can also be space-dependent as long as it respects the 
symmetry of Eq.~(\ref{eqVperiodic}).
A simple alternative to  Eq.~(\ref{Vexample})
would be to consider inclusions of regions with a certain refractive
index in a bulk with different properties, resulting in a piecewise
constant $V$.
This can be accomplished with laser-written waveguides~\cite{blomer2006nonlinear,szameit2006two}
or in typical photonic crystal fibers~\cite{ferrando2003spatial,ferrando2004vortex,brilland2006fabrication}.
An interesting possibility is that of hollow core fibers since the linear and nonlinear
properties can be tuned by appropriately filling them with a gas~\cite{benabid2005compact,trabold2014selective}.

We close this section by introducing some useful quantities.
The Hamiltonian density can be defined as
\begin{equation}
{\cal H}= \nabla \psi^* \cdot \nabla \psi + V(x,y) |\psi|^2 - \frac12 g |\psi|^4.
\end{equation}
We define its integral as the energy and notice that the norm of the wavefunction
is related to the power of the optical beam:
\begin{equation}
E = \int {\cal H} dx dy,\qquad  P= \int |\psi|^2 dxdy.
\end{equation}
It is straightforward  to check from Eq.~(\ref{eq1}) that  $dP/dz=dE/dz=0$.
In field theory, the energy is  the conserved quantity associated to time invariance. 
 In the optical case, the equivalent of a time translation is a spatial translation in the $z$-direction. 
 The conserved quantity is the generator of axial translations, namely, the $z$-component of the wave vector. 
 In optical waveguide theory, this quantity is used to classify the modes of a waveguide. 
 Thus, the propagation constant is the optical equivalent of energy in 
 (linear or nonlinear) waveguide theory.

\section{Eigenstates and stability}
\label{sec:eigenstates}

A propagation-invariant mode of Eq.~(\ref{eq1}) takes the form
$\psi=e^{-i \mu z} \varphi(x,y)$ for a real propagation constant $\mu$ with
\begin{equation}
\mu \varphi= - \nabla^2 \varphi + V({\bf x}) \varphi - g |\varphi|^2 \varphi.
\label{eq2}
\end{equation}

In the linear case ($P \to 0$ or $g=0$),  the structure of the lowest-lying 
 solutions of  Eq. (\ref{eq2}),
with the potential of Eq. (\ref{Vexample}) and the periodic boundary conditions of Eq.
(\ref{psiperiodic}), are readily obtained by means of Bloch and Wannier functions~\cite{kohn1959analytic}. We describe these functions in detail in appendix~\ref{sec:Bloch_lin}
 for the reader who is not acquainted with this formalism. 
 Each of the linear eigenfunctions gives rise to a solution of the 
 nonlinear problem Eq.~(\ref{eq2}).
 \begin{equation} 
\psi= e^{-i\mu_{\bf Q}z} \varphi^P_{\bf Q}({\bf x}),
\label{eq:nonlinBloch}
\end{equation}
 Here  $\varphi^P_{\bf Q}({\bf x})$  has the form of Eq. (\ref{blochlin}) and therefore is labelled by its pseudomomentum $\bf Q$ [with $N^2$ discretized possible values given in Eq.~(\ref{eq:Q})]. 
The $\varphi^P_{\bf Q}({\bf x}),$ can be 
 computed by propagating Eq. (\ref{eq1}) in imaginary
time.
Notice that $|\varphi^P_{\bf Q}({\bf x})|$  is periodic with period 1 (the physical period separating
the lattice sites) whereas $\varphi^P_{\bf Q}({\bf x})$, in general,
 has period $N$ (the  period associated
to the size of the computational domain).

\begin{figure}[h!]
\begin{center}
\includegraphics[width=\columnwidth]{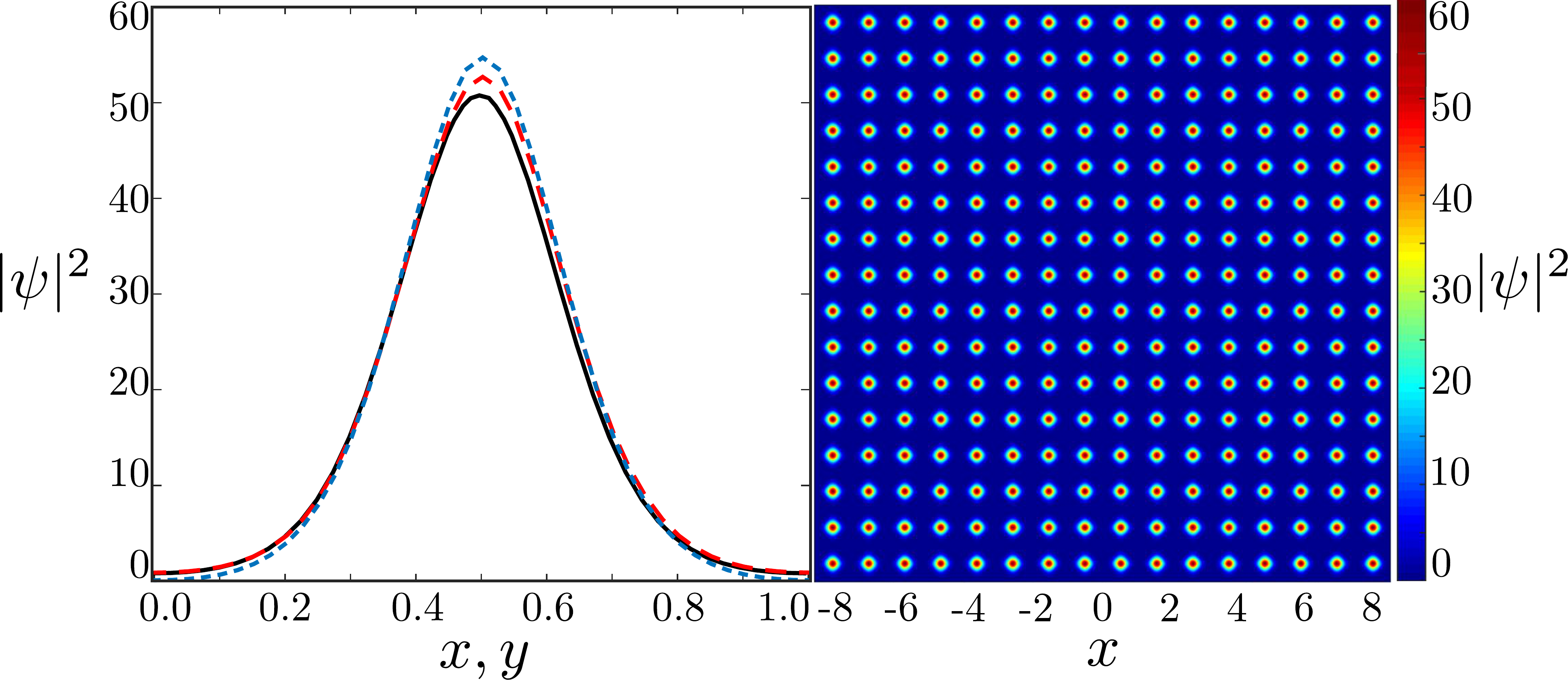}
\end{center}
\caption{Propagation-invariant solutions of the nonlinear equation with
$V_0=800$, $N=16$,
$P/N^2=5$.
On the left, $|\varphi^P_{\bf Q}(x,y=\frac12)|^2$ within a lattice cell for ${\bf Q}=(0,0)$ (solid black line),
${\bf Q}=(0,\pi)$ (dashed red line) and ${\bf Q}=(\pi,\pi)$ (dotted blue line).
Notice that for the first and third cases, it also corresponds to 
$|\varphi^P_{\bf Q}(x=\frac12,y)|^2$ due to the symmetry of the configuration.
On the right, a contour plot of $|\varphi^P_{\bf Q}(x,y)|^2$ with ${\bf Q}=(\pi,\pi)$ for the
whole domain. 
Notice that lattice sites have sides of length 1, which is the period of the refractive index and
of $|\varphi^P_{\bf Q}(x,y)|^2$. Periodic boundary conditions for $\psi$
are imposed at the boundaries of the computational domain of size $N\times N$. All units in figures are dimensionless - see Eq.~\eqref{eq1}. The dimensionless formalism is connected to photonic propagation in a periodic medium as discussed after Eq.~\eqref{parax1}. 
}
\label{fignew}
\end{figure}

In Fig. \ref{fignew}, we depict the light intensity profile for different values of ${\bf Q}$
for fixed $P$, $N$ and $V_0$. The nonlinearity is attractive and it can be seen that
the power is concentrated at the different lattice sites, without great qualitative differences
between the different values of ${\bf Q}$.

\begin{figure}[h!]
\begin{center}
\includegraphics[width=\columnwidth]{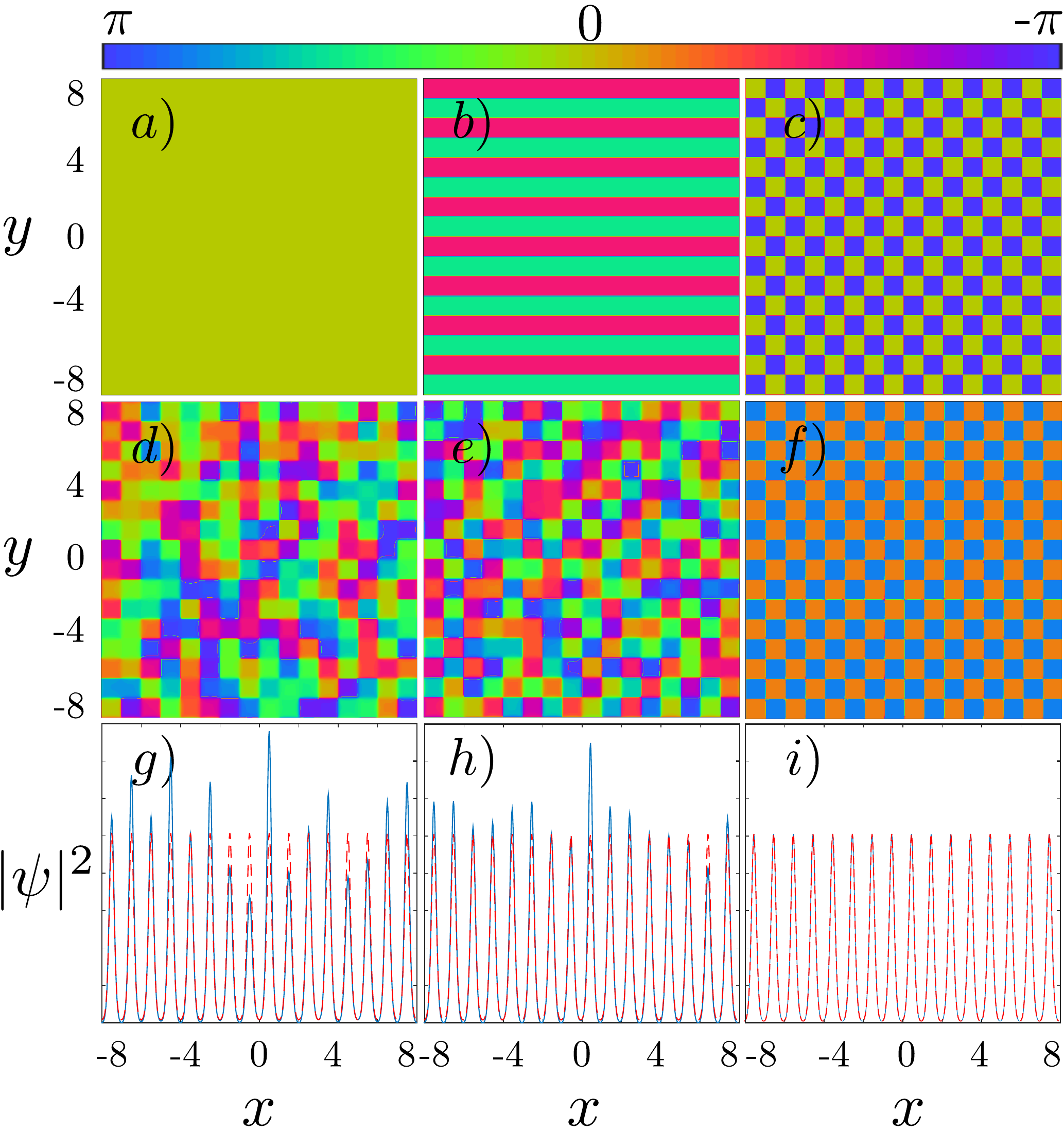}
\end{center}
\caption{
Evolution, computed by numerical integration of Eq. (\ref{eq1}), of three different  solutions 
of Eq. (\ref{eq2})
with $g=1$, $V_0=800$, $N=16$, $P/N^2=5$.
The unstaggered ${\bf Q}=(0,0)$ is represented in panels a) (phase distribution at $z=0$),
d) (phase distribution at $z=10$) and g) (the blue solid line represents
a one-dimensional cut $|\psi(x,\frac12)|^2(z=10)$ compared to the initial wavefunction 
depicted by the red dashed line $|\psi(x,\frac12)|^2(z=0)$).
With the same specifications, ${\bf Q}=(0,\pi)$ is represented in panels b), e), h) and
${\bf Q}=(\pi,\pi)$ in panels c), f), i).
 The figure shows that the  staggered configuration ${\bf Q}=(\pi,\pi)$ is the only stable one.}
\label{fig6}
\end{figure}

In fact, the difference between the modes of different ${\bf Q}$ is more clearly appreciated in
a plot for the phases, see first line of Fig. \ref{fig6}. In order to check the stability of the different
solutions, it is also convenient to depict the phase of the wave $\psi$. Upon evolution
(see second line of Fig. \ref{fig6}, computed by numerically integrating
Eq. (\ref{eq1}) with a standard beam propagation method), the phase pattern of the stable solution
remains ordered while the one for an unstable solution loses its initial disposition.
It turns out that for the focusing case $g=1$, the 
${\bf Q}=(\pi,\pi)$ solution, with its chess board-like phase pattern, is the stable
one. Conversely, in the defocusing case $g=-1$, the constant phase 
${\bf Q}=(0,0)$ solution is the stable one (not shown). 
We refer to the ${\bf Q}=(0,0)$ and ${\bf Q}=(\pi,\pi)$ configurations
 as unstaggered and staggered, respectively, borrowing the
condensed matter terminology for ferromagnetic and antiferromagnetic spin systems.

\begin{figure}[h!]
\begin{center}
\includegraphics[width=\columnwidth]{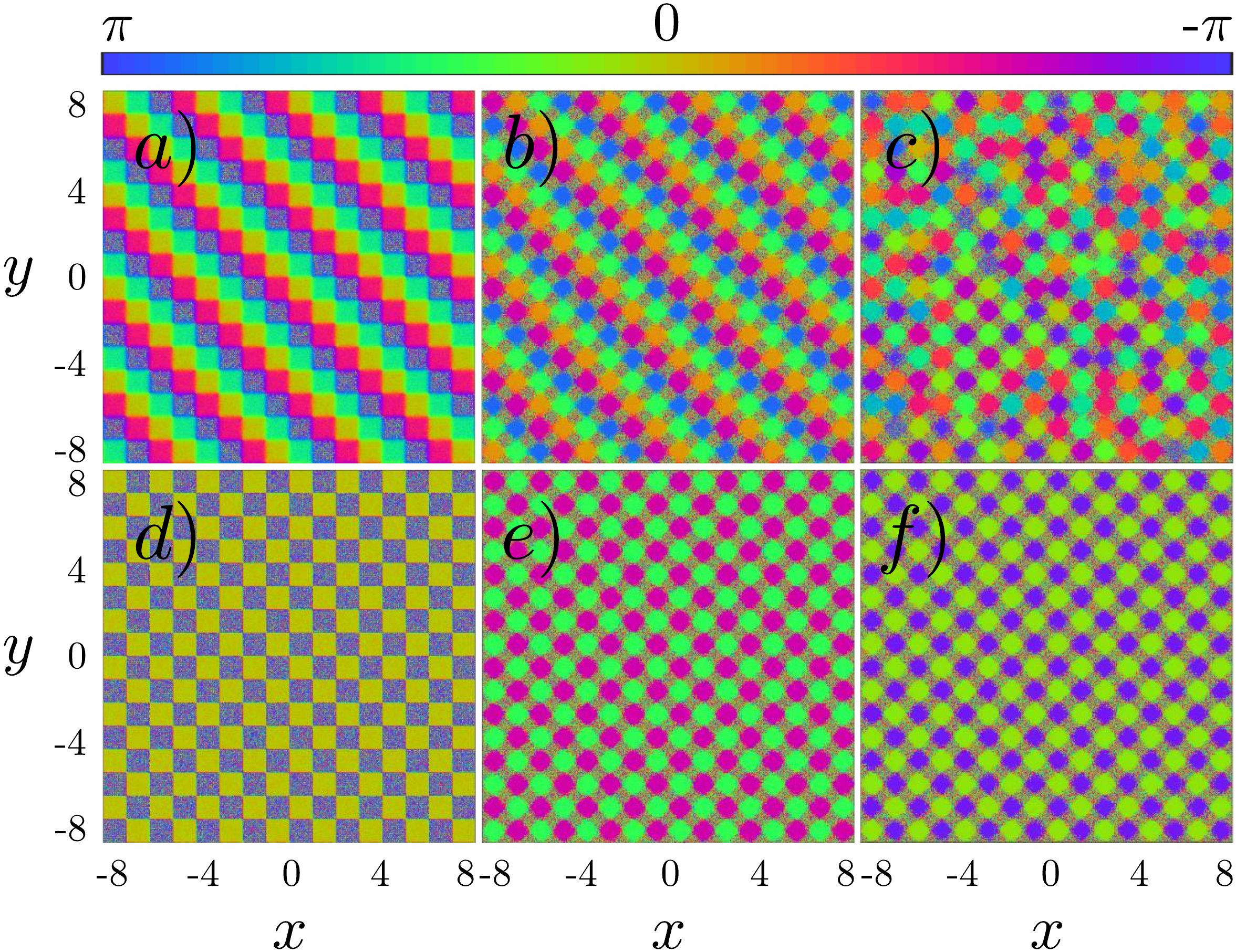}
\end{center}
\caption{Evolution of the phase for initial conditions with noise,
fixing $g=1$, $V_0=800$, $N=16$, $P/N^2=5$. Panels a), b) and c) correspond to
the ${\bf Q}=(\pi/2,\pi/2)$ case for $z=0$, $z=0.1$ and $z=4$ respectively.
Panels d), e) and f) correspond to ${\bf Q}=(\pi,\pi)$ for the same values of $z$.
The  noise is introduced by multiplying the initial phase at each point by a random
number from a normal distribution of amplitude 0.2.
It is visible that the phase structure gets destabilized in the ${\bf Q}=(\pi/2,\pi/2)$ case
whereas it remains stable in the staggered configuration.
}
\label{fig7}
\end{figure}

Some solutions, 
as for instance the
case of ${\bf Q}=(\pi/2,\pi/2)$  shown in the first column of Fig.~\ref{fig7}, can remain stable for large values
of $z$. For those cases, the simplest way to check that ${\bf Q}=(\pi,\pi)$ is the real
ground state is to introduce some random noise in the initial condition. What we see is that,
for small values of $z$, the noise in the phase is apparent in the regions where $|\psi|^2$
is small but the phase remains ordered otherwise. Under evolution, the initial phase pattern
breaks down, except for the staggered stable case.

The crux of this section is that focusing nonlinearity selects staggered phase patterns and
defocusing nonlinearity unstaggered ones. A few examples have been depicted but our numerical
simulations for many different cases suggest that this is a rather general conclusion. 
The effective description of Section \ref{sec:effective} will provide a natural
interpretation of this observation.

\section{Effective description}
\label{sec:effective}

The goal of this section is to introduce, borrowing definitions and methods from condensed matter
theory, a simple effective theory that allows us
to understand and predict general features of the propagation of light described by the
model of Eq. (\ref{eq1}). 
In particular, the description deals with any particular nonlinear solution $\varphi^P_{\bf Q}({\bf x})$ 
and
the fluctuations around it.
We  verify the conclusions by comparing them to  numerical 
computations.
The derivation is spelled out in detail in subsections \ref{sec:Bloch_nl}, \ref{free} and 
\ref{pert}.
The reader interested in the conclusions but not on the technicalities may skip
them.

\subsection{Nonlinear Bloch and Wannier functions}
\label{sec:Bloch_nl}

For a particular solution of Eq. (\ref{eq2}), $\varphi^P_{\bf Q}({\bf x})$, we
can define the following potential:
\begin{equation}
V_{sol} = V({\bf x}) - g |\varphi^P_{\bf Q}({\bf x})|^2.
\label{Vsol}
\end{equation}
Notice that $V_{sol}$ satisfies the  periodicity conditions of Eq. (\ref{eqVperiodic}).
A set of Bloch functions can be defined for $V_{sol}$,
 following the procedure of Appendix~\ref{sec:Bloch_lin}.
 They are called {\it nonlinear Bloch functions}
 and we denote them by $\varphi^{P,sol}_{\bf q}({\bf x})$.
  Nonlinear Bloch functions have appeared  in the literature in various contexts~\cite{Zhang2009Gap,Zhang2009Composition,Alexander2006Selftrapped,Carr2000Stationary,Carr2000Stationary2,Smirnov2007Observation}.  We consider here only the  $N^2$ of the lowest band. 
  Since $V_{sol}$ depends on $P$, the $\varphi^{P,sol}_{\bf q}({\bf x})$ also
  depend on the power, which 
  is a crucial difference with the linear case. Since the propagation constant 
  $\mu_{\bf Q}(P)$ also depends on $P$, the nonlinear Bloch functions depend on $\mu$. 
We  now introduce the
 nonlinear Wannier functions, 
\begin{equation}
W^P_{\bf R}({\bf x})=\frac{1}{N\sqrt{P}} \sum_{\bf q} e^{-i {\bf q}\cdot {\bf R}}
\varphi^{P,sol}_{\bf q}({\bf x}),
\end{equation}
where the sum runs over the $N^2$ Bloch momenta and ${\bf R}$  takes values corresponding
to the center of the $N^2$ lattice sites [see Eq.~(\ref{eq:posindexW})]. 
 By construction, $\varphi^P_{\bf Q}({\bf x})$ is itself a Bloch function of the $V_{sol}$ potential
and therefore, according to Eq.~(\ref{BlochWann}), it can be written as
\begin{equation}
\varphi^P_{\bf Q}({\bf x}) = \sum_{\bf R} c_{\bf R} W^P_{\bf R}({\bf x}),
\quad {\textrm {with}}\quad c_{\bf R}= \frac{\sqrt{P}}{N}e^{i {\bf R} \cdot {\bf Q}}.
\label{ci1}
\end{equation}
The  $W^P_{\bf R}({\bf x})$ are crucial for the  effective theory presented in the next section, 
since they constitute the most convenient basis to analyze fluctuations around
$\varphi^P_{\bf Q}({\bf x})$.

\subsection{The free energy}
\label{free}

The optical equivalent of the free energy is:
\begin{equation}
F=E-\mu\,P.
\label{Legendre}
\end{equation}
In terms of statistical physics, $P$ plays the role of the number of particles and 
$\mu$ corresponds to the chemical potential. 
The  relation  $P=P(\mu)$ for a family of nonlinear solutions,  or its inverse $\mu=\mu(P)$, 
can be understood as the equation of state in the equivalent system.
Along a family of nonlinear solutions, we find the  relations
$ \mu = \partial E/\partial P$ and
$P= -\partial F/\partial \mu$, linked by the Legendre transform (\ref{Legendre}).
This can be proved by considering a perturbation $\mu \to \mu + \delta \mu$,
$\varphi({\bf x}) \to \varphi({\bf x}) + \delta \varphi({\bf x})$. 
 Keeping only the leading terms in the perturbation, we can write
$\delta P = \int (\varphi^* \delta \varphi + \textrm{ c.c.}) d^2{\bf x}$ and
$\delta E = \int ( \delta\varphi [-\nabla^2 \varphi^* + V\varphi^* - 
g |\varphi|^2 \varphi^*] + \textrm{ c.c.}) d^2 {\bf x}$, where c.c. stands for 
complex conjugate.
Using Eq. (\ref{eq2}), it is straightforward to check that
$\delta E = \mu \delta P$.
The variation of the free energy within the family of solutions is given by 
$\delta F = \delta E - \mu \delta P - P\delta \mu = -P \delta \mu $. 

The free energy (\ref{Legendre}) can be expressed as an integral over the
entire domain $\Omega$
\begin{equation}
F= \int_\Omega d^2{\bf x} \left(\nabla \varphi^* \nabla \varphi + V({\bf x}) |\varphi|^2 - \frac{g}{2}|\varphi|^4 -
 \mu |\varphi|^2\right).
\label{free1}
\end{equation}
Integrating by parts and using Eq.~(\ref{eq2}), we find that its value for a  solution
of Eq. (\ref{eq2}) is
\begin{equation}
F=  \frac{g}{2} \int_\Omega d^2{\bf x} |\varphi|^4,
\label{onshellF}
\end{equation}
and  therefore it  is just characterized by the nonlinear interaction.

The next step is to integrate out the dynamics at each cell of the lattice in order to produce a discrete
model. With this goal, we use the expansion~(\ref{ci1}) in terms of nonlinear Wannier functions to rewrite the free energy~(\ref{free1})
in terms of the $N^2$ coefficients $c_{\bf R}$ associated to each cell.
For simplicity, we restrict ourselves to solutions with 
\begin{equation}
Q_x=Q_y = \frac{2\pi m}{N} \equiv q,\quad  m\in \left(-\frac{N}{2}+1 , \dots ,\frac{N}{2}  \right),
\label{Qm}
\end{equation}
 which present $x \leftrightarrow y$ symmetry. Then, keeping only the  on-site and nearest neighbor integrals of the Wannier functions,
we find
\begin{eqnarray}
F&\!=\!& \!\!\!\sum_{\bf R}\!\left(\!-t(\mu) \sum_{\nu=1}^4 c_{\bf R}^* c_{{\bf R} \!+\! {\bf n_\nu}} +
l(\mu) |c_{\bf R}|^2 - \frac{U(\mu)}{2} |c_{\bf R}|^4\! \right)\! \nonumber\\
\!&+& \!\!\!\sum_{\bf R}\!\left(\!-I(\mu)|c_{\bf R}|^2 \! \sum_{\nu=1}^4\! (c_{\bf R}^* c_{{\bf R} + {\bf n_\nu}}
\!+\!c_{{\bf R} + {\bf n_\nu}}^*  c_{\bf R})\!\!
\right)\! \!+ \!\dots
\label{discreteF}
\end{eqnarray}
where the ${\bf n_\nu}$ are the vectors connecting nearest neighbors, (1,0), (0,1), (-1,0), (0,-1)
and the dots represent higher order terms.
We have introduced the real quantities
\begin{eqnarray}
t(\mu)&=&- \int_\Omega W^{P *}_{\bf R}
\left(-\nabla^2 + V({\bf x}) - \mu\right) W^P_{\bf R+ n_1}d^2{\bf x},  \nonumber\\
l(\mu)&=&\int_\Omega W^{P *}_{\bf R}
\left(-\nabla^2 + V({\bf x}) - \mu\right) W^P_{\bf R}d^2{\bf x},  \nonumber\\
U(\mu)&=&g \int_\Omega |W^P_{\bf R}|^4 d^2{\bf x}, \nonumber\\
I(\mu)&=&g \int_\Omega |W^P_{\bf R}|^2 W_{\bf R}^{P *}W^P_{\bf R+ n_1}  d^2{\bf x}, 
\label{defcoeffs}
\end{eqnarray}
which are independent of the site ${\bf R}$ because of the translational symmetry of the
Wannier functions.  In this expression,
we have made explicit that $t$, $l$, $U$ and $I$ depend on the power $P$
and  ---due to the the equation of state $\mu=\mu(P)$--- on the propagation 
constant $\mu$, 
 because they are computed from the corresponding nonlinear Wannier functions, see
 Sec. \ref{sec:Bloch_nl}.

\subsection{Perturbations}
\label{pert}

In the language of the discretized theory, small perturbations around the
propagation-invariant 
solution of Eq. (\ref{ci1})
are described by promoting the $c_{\bf R}$ to $z$-dependent functions, in such a way
the uniform amplitude of the nonlinear Bloch function is substituted by a slowly varying ${\bf R}$-dependent envelope function
\begin{equation}
c_{\bf R}(z)= \Phi_{\bf R}(z) e^{i {\bf R}\cdot {\bf Q}}.
\label{cPhi}
\end{equation}

\begin{figure}[h!]
\begin{center}
\includegraphics[width=\columnwidth]{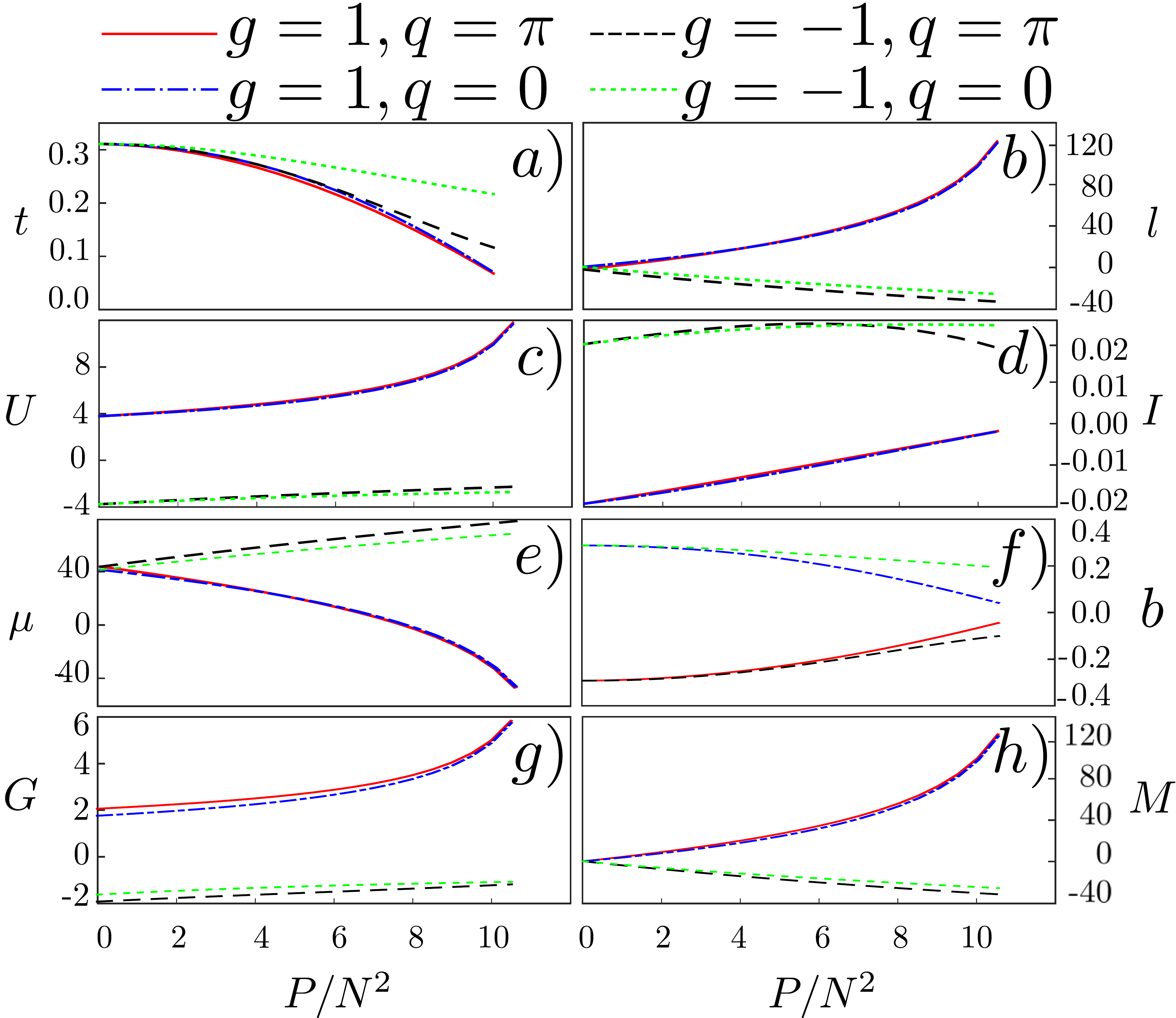}
\end{center}
\caption{In panels a)-d),  the values of the integrals of Eqs. (\ref{defcoeffs})
are shown for four cases with different
nonlinearity and different ${\bf Q}$, with
fixed $V_0=800$, $N=24$. Additionally, the low energy coefficients appearing in Eqs. (\ref{Fbar}) and (\ref{Fbar2}), for the effective free energy,
and in the field potential $(\ref{potential})$ 
are shown in panels f)-h) in terms of the normalized power $P/N^2$. In all cases ---except in the remarkable case f)--- the blue and red lines are practically overlapping
and in the plot are seen as a single solid line corresponding to $g=1$ with
$q=0$ and $q=\pi$, respectively. On the contrary, figure f) unveils that the sign of $b$ is positive for $q=0$ and negative for $q=\pi$. 
}
\label{fig5}
\end{figure}

The theory can be further simplified by going to a new  description in the continuum, valid for 
perturbations $\delta \Phi_{\bf R} $ with a typical length scale $\delta {\bf R} $ larger than the lattice spacing $a$. This is accomplished 
by transforming the summation over the ${\bf R}$ lattice sites of Eq. (\ref{discreteF}) into an integral,
with $ \Phi_{\bf R}(z) \to  \Phi({\bf x},z) $.
We write the resulting free energy as
\begin{eqnarray}
 F &=& \int_\Omega d^2{\bf x}  \Bigg[
b \nabla \Phi^* \nabla \Phi +
\left(M|\Phi|^2 - \frac{{\cal G}}{2}|\Phi|^4\right)+
\nonumber\\
&+&
I (\cos q) \Big(4|\Phi|^2 |\nabla \Phi|^2 + \Phi^{*2} (\nabla \Phi)^2 +
\Phi^{2} (\nabla \Phi^*)^2 
 \Big)+
\nonumber\\
&-& 2 i (\sin q) (t + 2 I |\Phi|^2)  \Phi^*(\partial_x \Phi + \partial_y \Phi) \Bigg],
\label{Fbar}
\end{eqnarray}
with
\begin{equation}
b=t \cos q,\quad 
{\cal G}=\frac{U}{2}+8I \,\cos q , \quad
M=l-4t \cos q.
\label{bh}
\end{equation}
In Fig. \ref{fig5}, we represent the values of $\mu$, and the coefficients of
Eqs. (\ref{defcoeffs}) and (\ref{bh}) as a function of $P$,
found numerically for a few examples. 

\subsection{Landau theory, spontaneous symmetry breaking and stability}
\label{sec:landau}

The expression of Eq. (\ref{Fbar}) 
is the sought effective description of the system that allows us to
determine the stability of  solutions and to study the nature of the ground state. 
In view of Eqs. (\ref{ci1}) and (\ref{cPhi}), the solutions with a definite pseudomomentum ${\bf Q}$
of the form $(\ref{blochlin})$ correspond in the effective description to a continuous field
with a uniform envelope $\Phi = \Phi_0$.

In order to simplify the discussion,  we 
deal with the leading order terms in $I(\mu)$ (compare in Fig. \ref{fig5} the scale for $I(\mu)$ in panel d)
to the rest of quantities) and, for reasons that will be clear next, neglect the last anisotropic term in Eq.(\ref{Fbar}). We then define the effective free energy 
describing the ground state and their low energy excitations as 
\begin{eqnarray}
\bar F & = & \textrm{sgn}(b) F = \int_\Omega d^2{\bf x}  \Bigg[ |b| \nabla \Phi^* \nabla \Phi \nonumber \\ 
& + & \textrm{sgn}(b) \left(M|\Phi|^2 - \frac{{\cal G}}{2}|\Phi|^4 \right) + \dots \Bigg] .  
\label{Fbar2}
\end{eqnarray} 
This change of sign does not alter the dynamics but ensures a positive kinetic 
term, as it is essential to define a proper ground, or vacuum, state \cite{coleman1988aspects}.

We now apply Landau theory, a mean-field approach used to
characterize phase transitions in condensed matter and
particle physics  \cite{chaikin2000principles}. 
We have to verify whether a constant configuration 
$\Phi = \Phi_0\neq 0$
can minimize (\ref{Fbar2}).
For $\sin q \neq 0$, that is not possible because long wavelength perturbations, 
e.g. $\Phi_0 \exp(i k x)$, produce a smaller $\bar F$ due to the anisotropic last term in Eq.(\ref{Fbar})
(this fact justifies the absence of the anisotropic term in Eq.(\ref{Fbar2})).
Therefore, we must set $q=0$, i.e.,
${\bf Q}=(0,0)$ (see panel (a) of Fig. \ref{fig2}) or $q=\pi$, i.e., ${\bf Q}=(\pi,\pi)$ (panel (d) of Fig. \ref{fig2}).

\begin{figure}[h!]
\begin{center}
\includegraphics[width=\columnwidth]{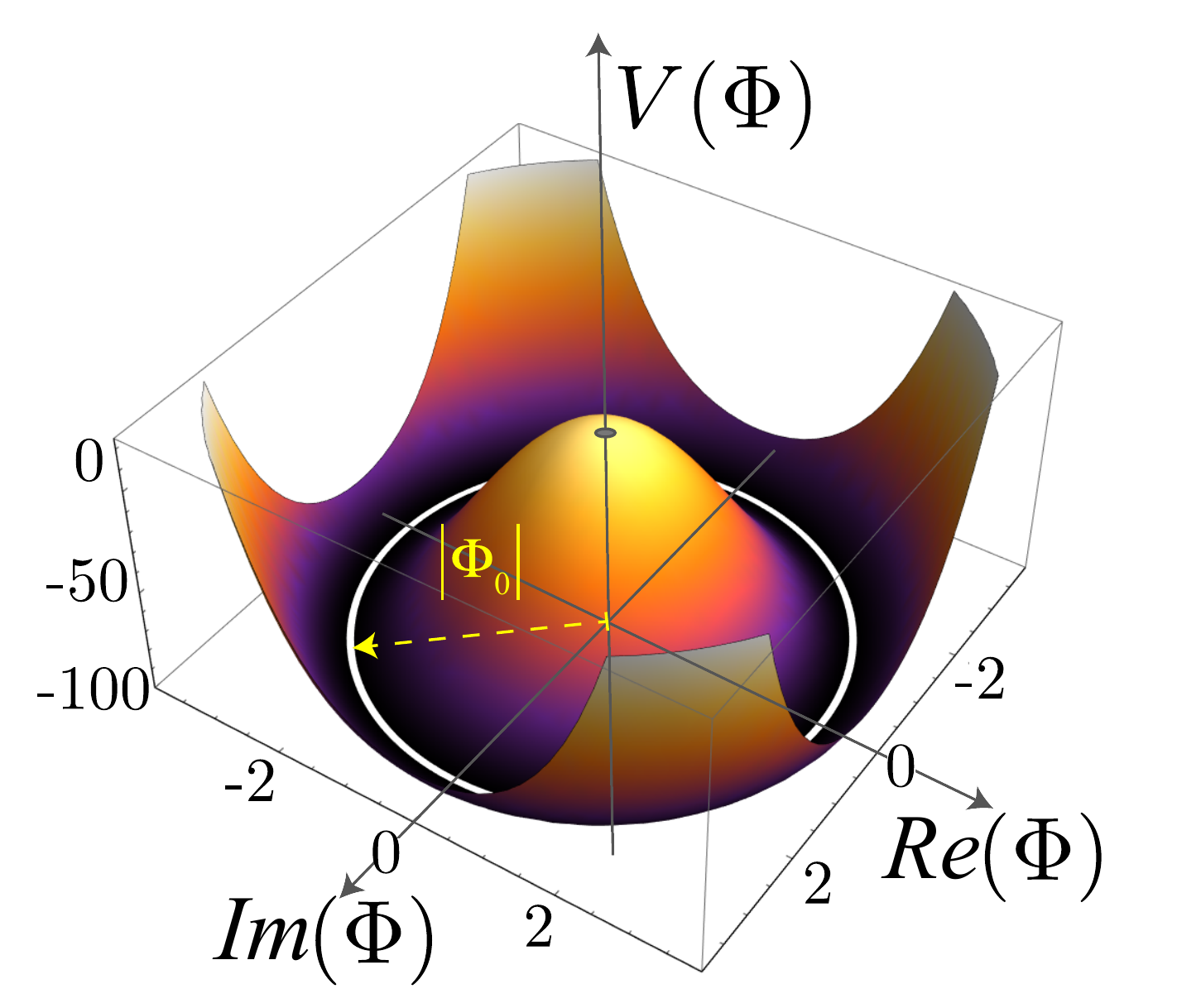}
\end{center}
\caption{
\emph{Calculated} effective field potential $V(\Phi)$ for a self-focusing  ($g=1$) and staggered 
($q=\pi$) 
solution in a lattice with
fixed $V_0=800$ and $N=24$ as in Fig.~\ref{fig5}. The power per site is $P/N^2=2$ so that $b=-0.3$, $M=11$ and $\mathcal{G}=2.2$.
The same Mexican hat shape is obtained in all cases (for these values of $V_0$ and $N$) provided  $g=1$ and $q=\pi$ or $g=-1$ and $q=0$.}
\label{fig5bis}
\end{figure}

From Eq. (\ref{Fbar2}) we see that, due to the fact that the kinetic term is now positive definite, a 
$\Phi=\Phi_0$
minimum of $\bar F$ exists if and only if it is a minimum of the effective field potential term
\begin{equation}
V(\Phi)=\textrm{sgn}(b) \left(M|\Phi|^2 - \frac{{\cal G}}{2}|\Phi|^4 \right).
\label{potential}
\end{equation}
 It is immediate to check that the nontrivial minimum condition  $\Phi = \Phi_0\neq 0$ is achieved when $(b\ M)<0$ 
and $(b\ {\cal G})< 0 $. 
The plots of Fig. \ref{fig5} show that this happens for the cases $g=1$, $q=\pi$ and
$g=-1$, $q=0$.  This is clearly visualized in Fig.~\ref{fig5bis}.

Thus, we find that for attractive nonlinearity, the ground state stable configuration
is the staggered one whereas for repulsive nonlinearity it is the unstaggered one, in
full agreement with the numerical computations of Section \ref{sec:eigenstates}.
The form of $V$ in Eq. (\ref{Vexample}) is not crucial for this result,
which  holds for generic types of periodic potentials. 
Heuristically, one may think of this result as a minimization of
the on-shell free energy (\ref{onshellF}), assuming that $|\varphi|^2$ is more spread out
in the unstaggered configuration, leading to a smaller $\int |\varphi|^4 d^2{\bf x}$ and 
that the opposite happens for the staggered solution (cf. Fig. \ref{fig1}).

In addition, the effective potential for a typical stable configuration, as shown in Fig.~\ref{fig5bis}, presents the 
paradigmatic Mexican hat shape.
This potential is
 characteristic of the spontaneous symmetry breaking mechanism appearing in
condensed matter, like in superfluids or superconductors \cite{Vollhardt2013}, and particle physics systems, like in the Higgs mechanism  \cite{coleman1988aspects}.
In situations described by this type of potential, the ground state is degenerate (white circle of radius $|\Phi_0|$ in Fig.~\ref{fig5bis}) 
presenting a continuous phase degeneracy: $\Phi_0(\alpha)=|\Phi_0| e^{i \alpha}$. Despite the original
free energy (\ref{Fbar2}) and  equation of motion are invariant under $U(1)$ 
phase transformations, the 
ground state $\Phi_0(\alpha)$ is not since it changes under a phase shift $\Phi_0(\alpha) \neq \Phi_0(\alpha+\theta)$. 
This fact has profound implications in the nature of the low energy spectrum of excitations, cf.
Section \ref{sec:goldstone}. 
In this approach, the stability of the ground state is equivalent to the stability of the original nonlinear Bloch solution,
represented by the uniform envelope $\Phi_0$ at a fixed value of  $P/N^2$.

The minimization of
the effective potential $V$ ---and thus $\bar F$--- requires
\begin{equation}
|\Phi_0|^2 = \frac{M}{{\cal G}}. 
\label{phi02}
\end{equation}
This quantity should be identified with the power in each
lattice site.
As a nontrivial crosscheck of the approximate effective model, we have verified
that the value of $M/{\cal G}$ computed from Eqs.
(\ref{defcoeffs}), (\ref{bh}) typically coincides with $P/N^2$ up to a deviation of a few percent.

This discussion based on Landau theory and on
the effective description of Eq. (\ref{Fbar}) provides a new point of
view for the interpretation of
previous theoretical \cite{kartashov2003two,alexander2006soliton} and experimental 
\cite{petrovic2003solitonic,neshev2004soliton,trager2006nonlinear}
results.
It is worth emphasizing that the dependence on $\mu$ ---and, therefore, on $P/N^2$--- of the coefficients of the
effective theory, which paves the way for the analysis \`a la Landau, is the result of
using the nonlinear Wannier functions in the modeling.

Even if the computations have been spelled out for a particular example,
we remark that the conclusion is rather robust, being applicable to large regions of the
space of parameters. There are some  limitations, on which we briefly comment now.
In the small $V_0$ limit, our approximations break down since the lattice structure 
becomes irrelevant and therefore the discussion loses its validity.
The same happens for large $P$, since, the wavefunction can self-focus and collapse in
each cell (for $g=1$) or overcome the periodic potential $V$ due to self-repulsion (for $g=-1$). 
Moreover, since this is a nonlinear phenomenon in nature, cases with
$P/N^2 \to 0$ have to be taken with care.

\section{Phase fluctuations as Nambu-Goldstone bosons}
\label{sec:goldstone}

The effective description of Sec. \ref{sec:effective} immediately yields another important dynamical
observation. As mentioned previously, we have a typical case of spontaneous  breaking of a continuous symmetry.
The theory (\ref{Fbar}) is invariant under a $U(1)$ phase rotation, but its ground state
is not. 
According to the well-known Goldstone's theorem \cite{coleman1988aspects}, such a symmetry breaking is accompanied by the existence of a massless 
Nambu-Goldstone boson related to the motion in field space along the generators of the
broken symmetry --- in our case of the $U(1)$ phase symmetry. This perturbation is always the one dominating the low energy 
dynamics \cite{Anderson1984,manohar1997effective}, since
it is easily excited, when compared to other degrees of freedom.
Well known examples of Nambu-Goldstone bosons are phonons
in superfluids or solids, magnons in ferromagnets \cite{chaikin2000principles}
 or the pions stemming from spontaneous chiral
symmetry breaking in quantum chromodynamics \cite{manohar1997effective}. 
Thus, on general grounds,
we expect to find long-range perturbations of the phase which propagate
as waves throughout the domain, that can be identified with Nambu-Goldstone bosons.

\begin{figure}[h!]
\begin{center}
\includegraphics[width=\columnwidth]{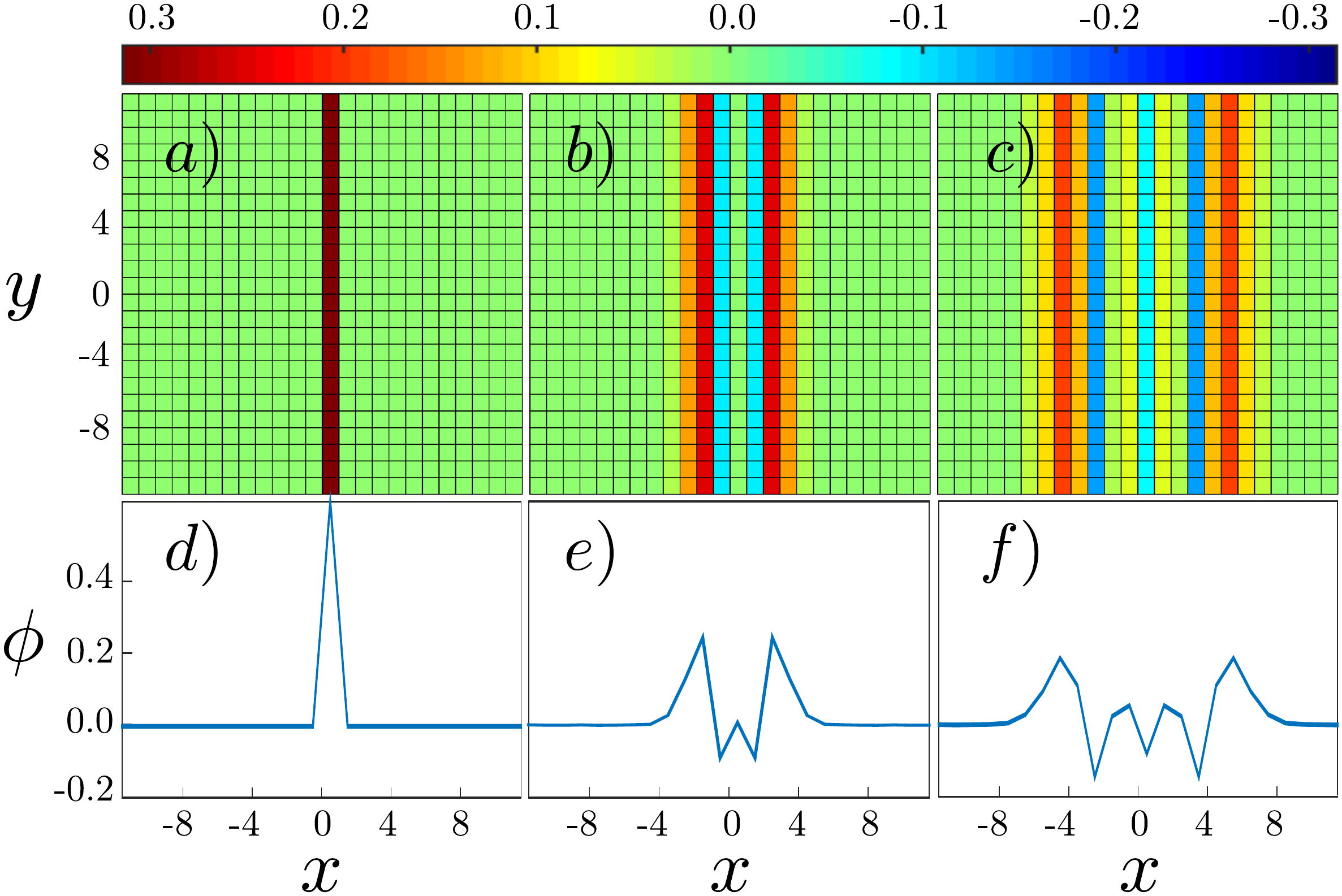}
\end{center}
\caption{Illustration of the propagation of the
excitation of the phase ($g=1$, $V_0=800$, $N=24$, $P/N^2=5$). 
The initial condition is like that of panel c) of Fig. \ref{fig6}, multiplying
by $e^{i \pi/5}$ the wavefunction for a column of lattice sites.
Panels a)-c) depict the phase distribution for three values of $z$.
Panels d)-f) are $y=0$ cuts of the same quantity.
Notice that in panel a) the strength of the perturbation ($\pi/5$) is higher than the
maximum of the
 colormap scale, a convention chosen to improve visualization in
 panels b) and c).
}
\label{fig8}
\end{figure}

In order to illustrate the process, we take a stable  solution and introduce a 
phase shift in some of the lattice cells, see Fig. \ref{fig8}. 
In the example, we take a staggered solution with focusing nonlinearity and, at $z=0$, we
multiply the $\psi$ of the cells of the central column by $e^{i \pi/5}$. We then
integrate numerically Eq. (\ref{eq1})  and depict how the perturbation of the phase propagates within
the lattice, generating a wave moving leftwards and another one moving rightwards.
We defer to appendix \ref{appendix} a precise definition of the phase perturbation plotted in the figure.
The modulus of $\psi$ and, with it, the power per lattice cell, also gets perturbed. However, the size
of its perturbation decreases as the $P/N^2$ of the background solution increases and is subdominant
with respect to the phase oscillation. In this sense, we refer to the excitation as a perturbation of the
phase.

From a simulation like the one of Fig. \ref{fig8}, we can infer the velocity of the wave
by, e.g., following the position of the first peak moving leftwards and fitting its position in
terms of $z$ to a line. For fixed $g$ and $V_0$, we find that it does depend very mildly on
the strength of the perturbation and on $N$. On the other hand, it does change significantly
with the value of $P/N^2$ of the background solution, 
stressing the importance of the nonlinear effects and of the usage
of the nonlinear Wannier functions. 

\begin{figure}[h!]
\begin{center}
\includegraphics[width=.8\columnwidth]{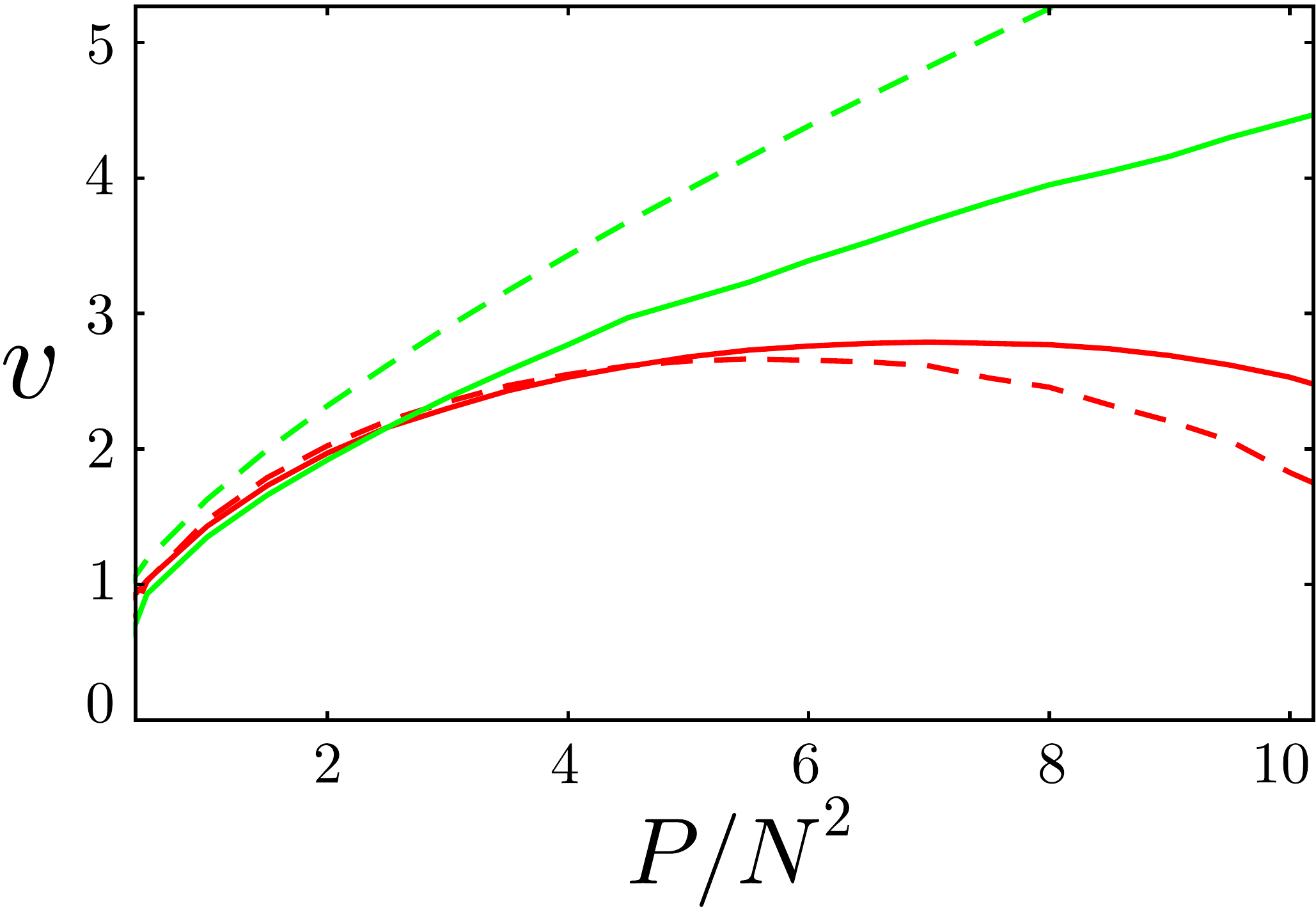}
\end{center}
\caption{Comparison of the wave velocity obtained from
repeated numerical integration (solid line)
of Eq. (\ref{eq1}) and the one predicted by the effective model (dashed line) in Eq.(\ref{vmodel}) as a function
of the power per lattice cell.
The green upper curves correspond to the defocusing unstaggered case and 
the red lower curves correspond to the focusing staggered case.
Notice that the horizontal axis starts at $P/N^2=0.5$ because when this quantity tends to zero
the excitation loses its Nambu-Goldstone boson character.
}
\label{fig9}
\end{figure}

We now discuss this dynamics in the framework of the effective model of Sec. 
\ref{sec:effective}.
Introducing a small perturbation  
\begin{equation}
\Phi=\Phi_0(1 + \gamma ({\bf x},z)) \exp(i \alpha ({\bf x},z)) \quad (\gamma,\alpha \in \mathbb{R})
\label{Phi}
\end{equation}
in the equation of motion $i\partial \Phi / \partial z = \delta {\cal F} / \delta \Phi^* -
\nabla (\delta {\cal F} / \delta \nabla \Phi^*) $, where ${\cal F}$ is the integrand
in Eq. (\ref{Fbar}) and assuming $\sin q=0$, we find
\begin{eqnarray}
\partial_z \gamma &=&- \cos q(t + 2 I \Phi_0^2)\nabla^2 \alpha, \nonumber \\
\partial_z \alpha&=&2 M \gamma - \cos q(t + 6 I \Phi_0^2) \nabla^2 \gamma.
\end{eqnarray}
By taking the axial derivative in the second equation, substituting the value of $\partial_z \gamma$ according to the first
equation and keeping only the leading order terms in the derivative expansion, one finds the
equation fulfilled by the phase excitation $\alpha$ at long distances or, equivalently, at low energies:
\begin{equation}
\nabla^2 \alpha - \frac{1}{v^2}\partial_z^2 \alpha =0.
\label{NGeq}
\end{equation}
This equation is the wave equation in two spatial and one time dimensions (where $z$ plays the role of the temporal coordinate). In the language
of condensed matter and particle physics, this equation corresponds to the massless excitation associated to the phase,
which is the Nambu-Goldstone boson associated to the symmetry broken by the ground state, namely, the $U(1)$  phase symmetry as depicted in the 
Mexican hat potential in Fig.~\ref{fig5bis}. In addition, our calculation predicts an explicit expression for the phase velocity
of the Nambu-Goldstone phase excitation:
\begin{equation}
v=\sqrt{-2 M (\cos q) (t + 2 I |\Phi_0|^2)}. 
\label{vmodel}
\end{equation}
It is important to remark that
$M$, $t$, $I$ and $ |\Phi_0|$ are all nontrivial functions of $P^2/N$, as shown in Fig.~\ref{fig5}.
For the focusing unstaggered and defocusing staggered cases, which we found to be unstable
in Sec. \ref{sec:effective}, this $v$ is imaginary, providing a new evidence of instability.
For the focusing staggered and defocusing unstaggered cases, Eq. (\ref{vmodel})
gives an estimate that can be compared with the results of direct numerical
integration of Eq. (\ref{eq1}). This is done in Fig. \ref{fig9}. 

Equations (\ref{Phi})-(\ref{vmodel}) are the central result of this paper, since they
allow us to define a propagation mode from the effective theory that matches the
direct numerical computations. Thus, they 
 confirm that the discussed phase perturbations are, indeed, Nambu-Goldstone bosons.
 They also show that the velocity of the perturbation depends on the intensity of the light beam.

The wave equation (\ref{NGeq}) leads to a linear dispersion relation between $\omega$ ---the conjugate variable of
the evolution variable $z$--- and the spatial wavevector $\mathbf{k}$: $\omega \sim |\mathbf{k}|$. This is a
remarkable fact since the dispersion relation associated to the original free energy (\ref{Fbar2}) for the envelope field $\Phi$
is quadratic: $\omega \sim |\mathbf{k}|^{2}$. In terms of spatio-temporal symmetries, the Galilean invariance of the original
description in terms of a generalized nonlinear Schr\"odinger equation turns into the Lorentzian invariance of the wave equation
(\ref{NGeq}) for the $\alpha$ phase field. This is a known effect in condensed matter physics associated to the spontaneous 
breaking of phase symmetry, as in superfluidity and superconductivity \cite{Anderson1984, chaikin2000principles, Vollhardt2013}. A complete analysis
in the language of effective field theories \cite{manohar1997effective}, but in terms of spontaneous symmetry breaking in non-relativistic systems provides the
same answer~\cite{GREITER1989a,Leutwyler1994a,Escobedo2010a}.

The plot of Fig. \ref{fig9} shows that the effective model correctly captures  the qualitative 
features of the Nambu-Goldstone boson phase excitation. For small $P/N^2$, the velocity grows 
in an approximately linear fashion. This can be understood heuristically by noticing that,
 in the first term
of Eq. (\ref{Fbar2}), if we split $\Phi$ in modulus and phase, the kinetic term for the phase is
multiplied by $\Phi_0^2 \approx P/N^2$. If we integrate in $z$ to obtain the action, $dz$ is 
multiplied by $P/N^2$, resulting in the mentioned linear dependence. For focusing
nonlinearity, the velocity reaches a maximum and then decreases. This happens because the
light distribution within each site gets more and more spatially confined, reducing the 
interaction with the neighboring sites. For larger values of $P$, the wavefunction would self-focus and
collapse within each site (we notice that in the present dimensionless convention the critical power of the
Townes profile is 11.7). One can also appreciate that the quantitative match of the model with the 
numerics is better in the focusing case. The interpretation is that for defocusing nonlinearity, the
light distribution is more spread out and the notion of lattice discretization becomes less clear,
 limiting the quantitative validity of the approximation leading to
Eq. (\ref{discreteF}) and thus to all expressions derived from it, including the one for $v$. Finally, let us remark that, as it happens in typical cases of
particle or condensed matter physics, the quantitative precision of the approximation could get
better if additional terms are included in the expansion leading to the effective theory \cite{manohar1997effective,GREITER1989a,Leutwyler1994a}.

\begin{figure}[h!]
\begin{center}
\includegraphics[width=\columnwidth]{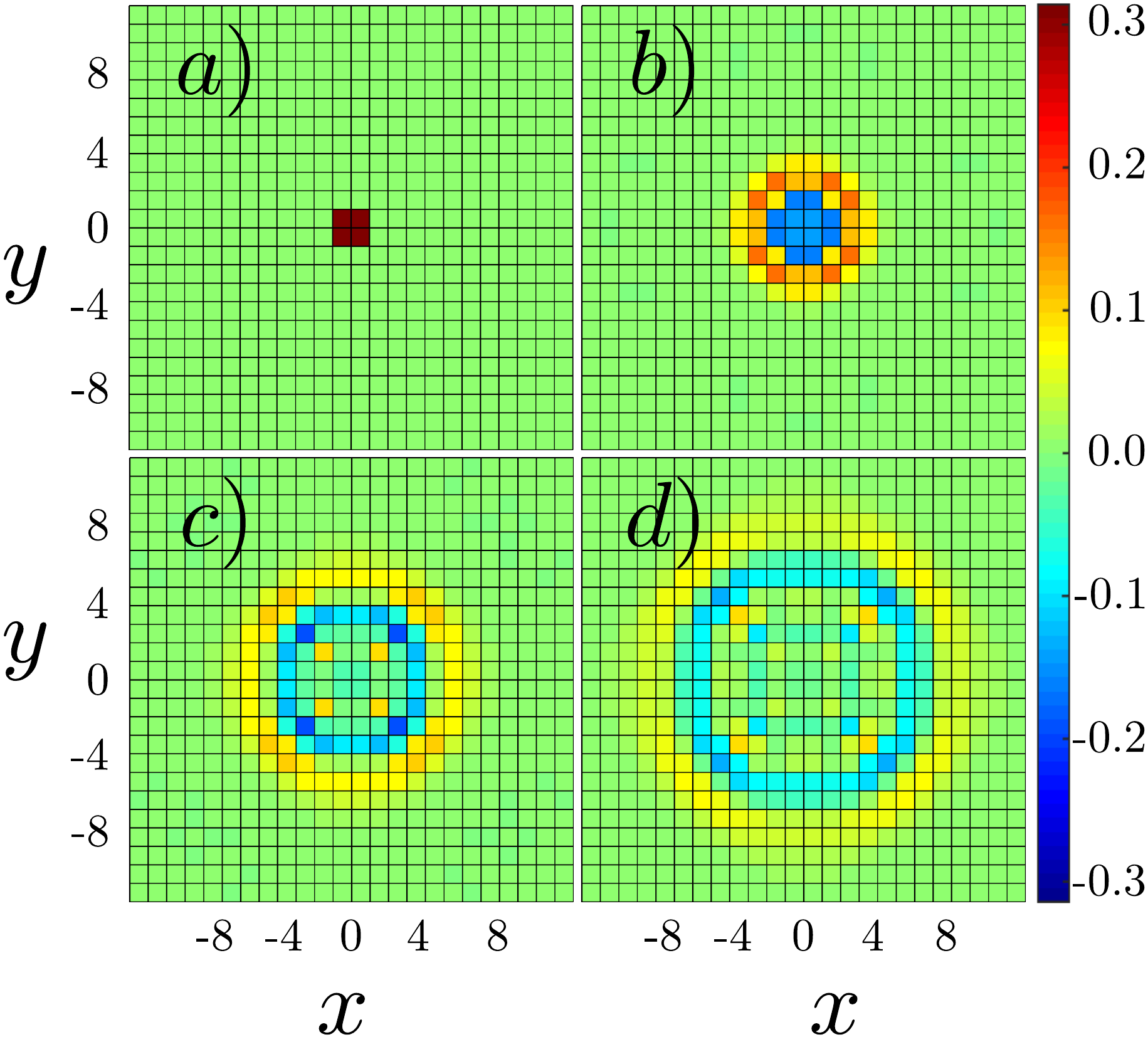}
\end{center}
\caption{Propagation of a circular wave. Initial conditions are as in Fig.
\ref{fig8}, except that the $\pi/5$ initial phase perturbation is performed
on the four central cells of the lattice.
Notice that in panel a) the strength of the perturbation ($\pi/5$) is higher than the
maximum of the
 colormap scale.
}
\label{fig10}
\end{figure}

Of course, the profile of the wave depends on the symmetry of the initial conditions.
In Fig. \ref{fig10}, we plot the result of initially perturbing the phase of four adjacent cells, resulting
in a  circular-shaped wave.

\section{Tunable metawaveguides for phase excitations}
\label{sec:finite}

In the previous sections, we have used periodic boundary conditions for the wavefunction, 
Eq. (\ref{psiperiodic}). Thus we assume that the entire periodic structure is illuminated. This was done for mathematical convenience, in order to 
introduce and utilize Bloch and Wannier functions, as in condensed matter theory.
As far as phase excitations are concerned, the effective ground state defined by the uniform envelope $|\Phi_0|$ of these nonlinear Bloch 
functions, which extend over the entire domain, acts as an effective optical material for phase waves.
For different values of power, the nonlinear Bloch function behaves as a
different metamaterial ---made of light--- for the Nambu-Goldstone phase waves propagating on top of it, as unveiled in Fig.~\ref{fig9}.

In order to make contact with realistic situations, we need to discuss what happens
for finite materials and spatially finite solutions. The expectation, that we will check in this section by
numerical computations, is that most of the conclusions can be directly generalized to
this case. Notice that, even if the Wannier functions were defined making use of the periodicity
conditions, they are essentially localized functions (see Fig. \ref{fig4} for illustration).
Thus, the effective description of Eq. (\ref{Fbar}) holds essentially unchanged, at least far
from the borders of the structure, and the predictions about stability and the presence of a Nambu-Goldstone boson excitation
apply. Certainly, the larger the finite  lattice
is, the better the approximation becomes.

Conceptually, since Bloch solutions act as a metamaterial for phase waves, finite size
solutions embedded in a photonic lattice can be used as effective waveguides to control phase excitations, as pictorially depicted in Fig.~\ref{fig0}. 
 Since
at low energies the Poynting vector is approximately given by $\mathbf{S}\sim |\Phi_0|\nabla \alpha$, a full optical control of the electromagnetic flux itself can be
also achieved using these metawaveguides.

\begin{figure}[h!]
\begin{center}
\includegraphics[width=\columnwidth]{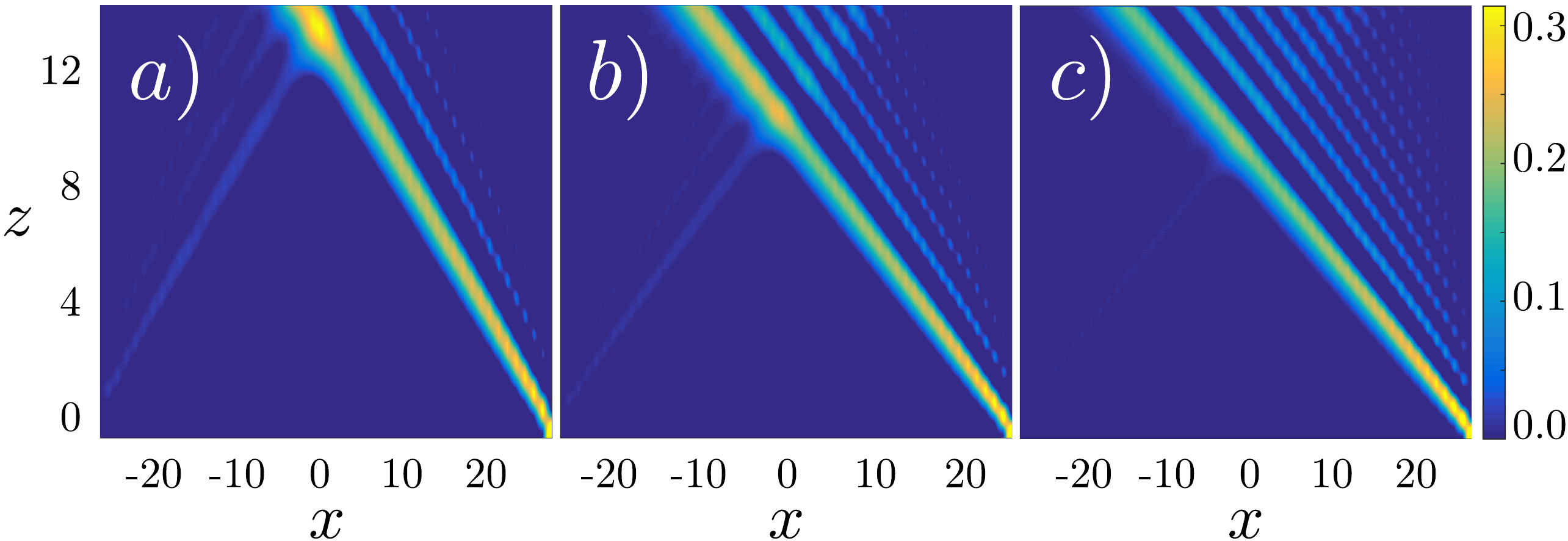}
\end{center}
\caption{One-dimensional cuts of the phase perturbation
$\phi (x, y=\frac12 ,z)$ moving leftwards,
initiated by initial conditions with a $\pi/5$ phase shift in the right-most column.
In this example $g=1$ and $V_0=800$. The different panels 
correspond to different values of $P/N^2$, namely
 $P/N^2=2$ in panel a), $P/N^2=4$ in panel b) and $P/N^2=6$ in panel c).
The color scale does not cover the full range of the perturbation $\phi$, but it has
been chosen to optimize the visibility of the plots without missing essential information. 
}
\label{fig11}
\end{figure}

In Fig. \ref{fig11}, we present some examples of propagation in a finite structure.
We define a rectangle of $52 \times 8$ unit cells where $V$ takes the form of Eq. (\ref{Vexample}).
Outside the rectangle, we fix $V=V_0$, hindering the diffusion of light to that region.
With the method of propagation in imaginary time,
we find the stable ground state of the system, which is, approximately, of the form
(\ref{blochlin}). Then, we introduce, at the border of the rectangle,
 a perturbation of the phase, similar to
 the one depicted in
  Fig. \ref{fig8} and compute the evolution of the system.
 As expected, the perturbation of the phase propagates in a similar fashion to the fully periodic case, apart
 from the fact that, since the perturbation starts from the edge, it moves unidirectionally.
 For a given $g$ and $P/N^2$, we can infer the velocity which turns out to coincide, to the few percent
 level, with that depicted in Fig.\ref{fig9}.

The plots of Fig.~\ref{fig11} show explicitly that velocity of the Nambu-Goldstone phase excitation
is different when perturbing
solutions with different values of $P/N^2$, emphasizing the nonlinear character of the
 dynamics. More subtly, within each of the plots, small dispersive
effects can be appreciated, coming from mild dependences of the velocity on the strength and
the wavelength of the perturbation.
Curiously, another wave appears moving rightwards from the left border, because the initial
condition is not an exact eigenstate. As expected from the discussion of 
Section \ref{sec:goldstone}, its velocity is very similar to the
one of the main perturbation.

\begin{figure}[h!]
\begin{center}
\includegraphics[width=\columnwidth]{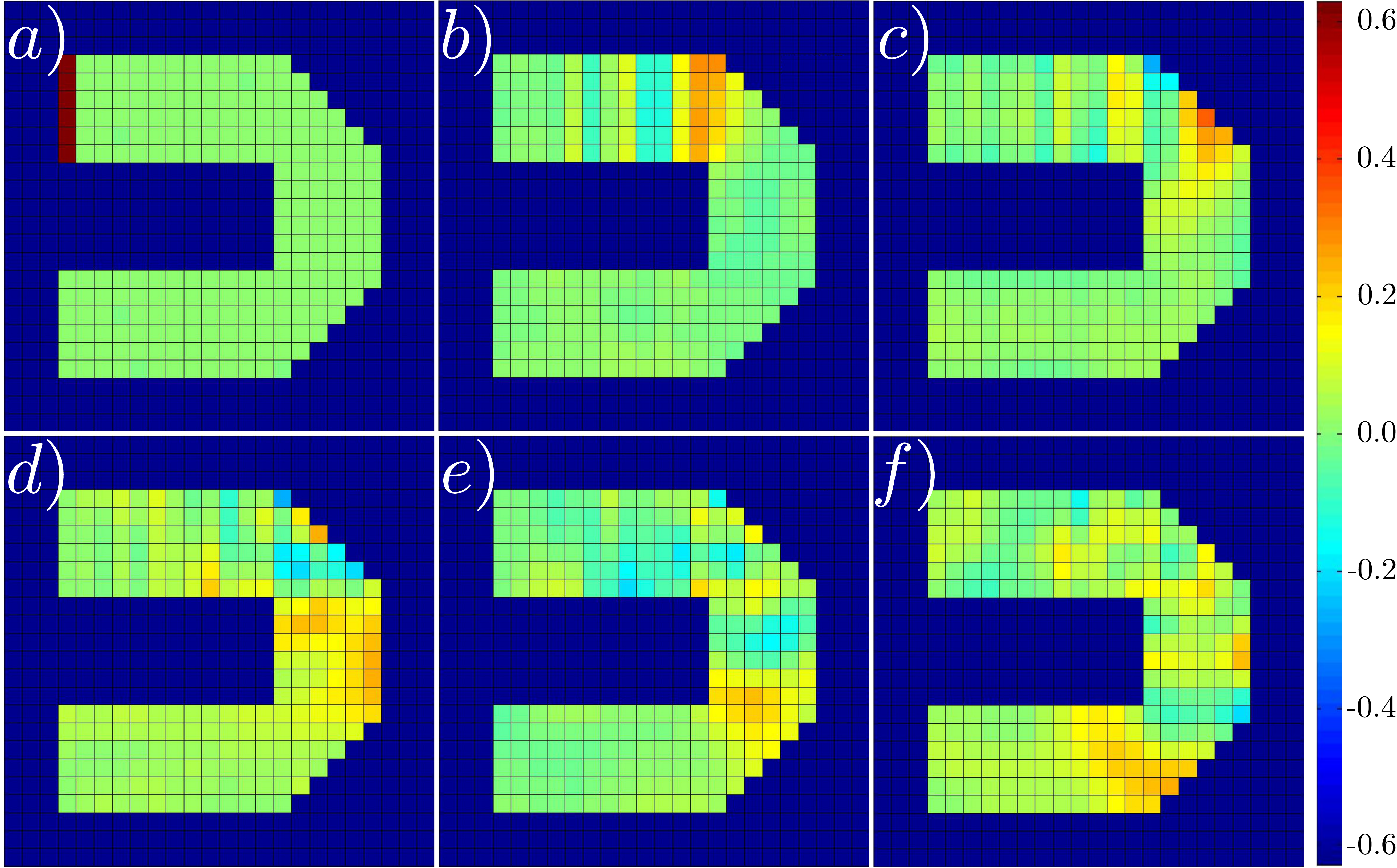}
\end{center}
\caption{Bounce of a Nambu-Goldstone phase wave due to reflection
in the edge of the structure, resulting in a shift of the propagation direction.
The perturbation of the phase is depicted in the lattice sites where
Eq.~(\ref{Vexample}) holds. For the rest, where $V=V_0$, the modulus
$|\psi|^2$ is nearly vanishing and we just plot them in a uniform color to help 
visualization.
In this example $g=1$, $V_0=800$ and $P/N^2=5$.
 Panels a)-f) correspond, respectively to $z=0,4.5,6,7.5,9,10.5$.
}
\label{fig12}
\end{figure}

It is interesting to discuss the fate of a perturbation that reaches the boundary of the 
finite sample. It turns out that the edges act, at least qualitatively, as a reflecting surface.
In this way, this structure becomes a paradigmatic case of a metawaveguide ---formed by light itself--- discussed previously.
It is tunable because the velocity of the phase perturbations can be 
controlled by changing the intensity of the light within the photonic lattice,
see Eq. (\ref{vmodel}) and Figs. \ref{fig5}, \ref{fig9} and \ref{fig11}.
There are a number of possibilities when considering non-trivial geometries as, e.g., 
having the potential of Eq. (\ref{Vexample}) for some sites and constant $V=V_0$ for the rest.
In Fig.~\ref{fig12}, we present an example of a U-shaped metawaveguide, and it can be seen
how the perturbation turns around due to its interaction with the borders. 
This opens the possibility of nontrivial manipulation of light through nonlinear effects 
and of guiding the phase perturbations of Section \ref{sec:goldstone}. 

\section{Discussion and outlook}
\label{sec:discussion}

An ever-increasing experimental control of the nonlinear propagation of light in nontrivial
media has led in recent years
 to the observation in optical setups of many qualitatively new phenomena.
 In particular,
 this has sparked remarkable interest in finding conceptual and quantitative connections
with condensed matter physics. 

We have shown that methods borrowed from
condensed matter physics can be useful for the optical community.  
We have exploited the nonlinear character of the Schr\"odinger equation~(\ref{eq1}) to set the bridge between nonlinear optics and condensed matter in a quantitative manner. Expanding the solutions of  Eq.~(\ref{eq1}) in terms of nonlinear Wannier functions, we have been able to obtain an effective description, Eq.~(\ref{Fbar}), of the dynamics of light propagating within a nonlinear material with a transversally symmetric potential. In this context, we used Landau theory to formally establish  the existence of the paradigmatic Mexican hat 
potential, archetypal of the spontaneous symmetry breaking mechanism appearing in condensed
matter and particle physics. This connection has been posed 
not only qualitatively, but we have computed 
the effective potential in terms of the coefficients of the problem, 
showing the key role of the nonlinearity. In light of this effective potential we have been able to determine the stability criteria for the solutions both in the focusing and the defocusing cases. 

Two crucial outcomes arise naturally from this point of view: first, the existence of a Nambu-Goldstone boson, which is identified with the phase excitations. We have presented the equation fulfilled by the phase excitation, 
Eq.~(\ref{NGeq}). Again, this is quantitatively evaluated in terms of the parameters of the problem, agreeing with direct numerical simulations (see Fig.~\ref{fig9}, where
 the predicted and numerically calculated phase velocities  are compared). 
 The Nambu-Goldstone boson occurs both for self-defocusing and focusing
 non-negligible nonlinearity,  in the latter case 
 well below the limit for self-focusing and collapse within each lattice cell. 
The second outcome is the concept of  a metawaveguide of light, discussed in  Section \ref{sec:finite}: 
the phase excitations identified as Nambu-Goldstone bosons can be guided and controlled by the underlying nonlinear localized light structure. We have presented numerical examples of how to 
guide such a perturbation (cf Fig.~\ref{fig12}).

For an experimental observation, and since the main perturbation is in the phase and not
in the intensity, detection based on interferometry would be needed, and it could constitute a
notable challenge. This requires interference with a reference beam in a similar way 
as it is conventionally done for vortex structures \cite{neshev,keyrole}

We have not discussed topological defects like vortices 
(see e.g. \cite{terhalle2008observation} for related work) and domain walls, that are also 
immediate consequences of the described spontaneous symmetry breaking. 
Its theoretical and/or experimental analysis is beyond the scope of the present work.
Certainly, we have limited the discussion to a particular family of examples. 
It would be of interest to extend it to more general cases as, e.g.,
lattices with different symmetries \cite{desyatnikov2006two},
 lattices with modulation \cite{garanovich2012light}, anisotropic
 lattices \cite{terhalle2007anisotropic} or with the inclusion of gain and loss
 in parity-time symmetric systems \cite{regensburger2012parity}.

We have also mentioned that our discussion might be applicable to Bose-Einstein 
condensates. It is worth mentioning some works that analyze stability in
related setups, including one-dimensional \cite{menotti2003superfluid,fallani2004observation,barontini2007dynamical,zhang2013superfluidity} 
and two-dimensional cases
\cite{chen2010stability,xu2013stability}. Whether the approach advocated here can
provide new insights on these or other similar examples is an open question for the future.
It is worth mentioning that Nambu-Goldstone modes have been discussed in the context
of atomic condensates in relation to the breaking of non-abelian symmetry groups \cite{uchino}.

Finally, our work might pave the way for new connections between classical results
of condensed matter theory and the propagation of light within properly tuned media.
Notice, for instance, that symmetry and symmetry breaking have been 
discussed in the context of cavity
quantum electrodynamics \cite{fam,chilingaryan}.
Another possibility could be  to devise an optical analogue of the Heisenberg
model used in statistical physics to model ferromagnetism. That setup could mimic the
subtle structure of phases and phase transitions in two-dimensional
systems \cite{mermin1966absence,kosterlitz1973ordering}, which is a manifestation
of the rich and sometimes counterintuitive dynamics of many body systems.
It can also provide a useful framework to analyze the role of light in recent experiments showing superfluidity
in polariton condensates \cite{Lerario2017a}. Our work also stresses the role played 
by phase excitations of nonlinear Bloch modes. 
Recent experiments in quantum communication systems use  the phase of light to encode information,
 thus showing its advantages over conventional encoding of the information in the
  amplitude \cite{Cao2017}. We expect that the phase mode discussed here is a good 
  candidate for encoding and guiding information in this context. 
The concept of a metawaveguide of light for the control of phase waves fits nicely within this scenario.

Our main claim is that a \emph{photonic condensed matter } formalism is the perfect theoretical
framework to analyze the exciting analogies between condensed matter (and particle physics) and photonic systems.

\appendix

\section{Bloch and Wannier functions in the linear case}
\label{sec:Bloch_lin}

\begin{figure}[h!]
\begin{center}
\includegraphics[width=\columnwidth]{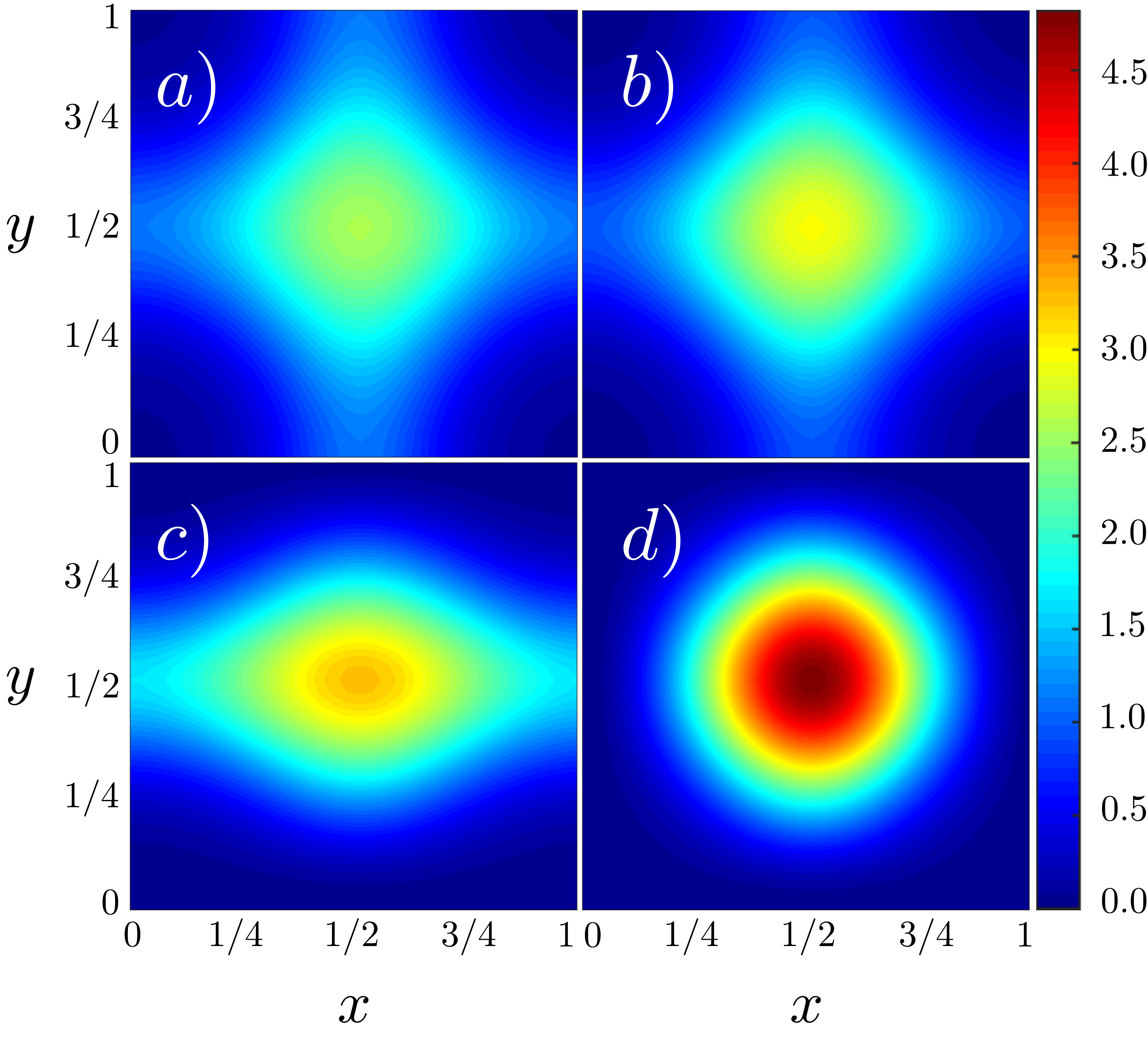}
\end{center}
\caption{Representation of $|u_{\bf Q}(x,y)|^2$ for $V_0=100$ and
 different values of
${\bf Q}$. Panel (a): $Q_x=Q_y=0$. Panel (b): $Q_x=Q_y=\pi/2$. Panel (c) $Q_x=0$, $Q_y=\pi$.
Panel (d): $Q_x=Q_y=\pi$. 
}
\label{fig1}
\end{figure}

\begin{figure}[h!]
\begin{center}
\includegraphics[width=\columnwidth]{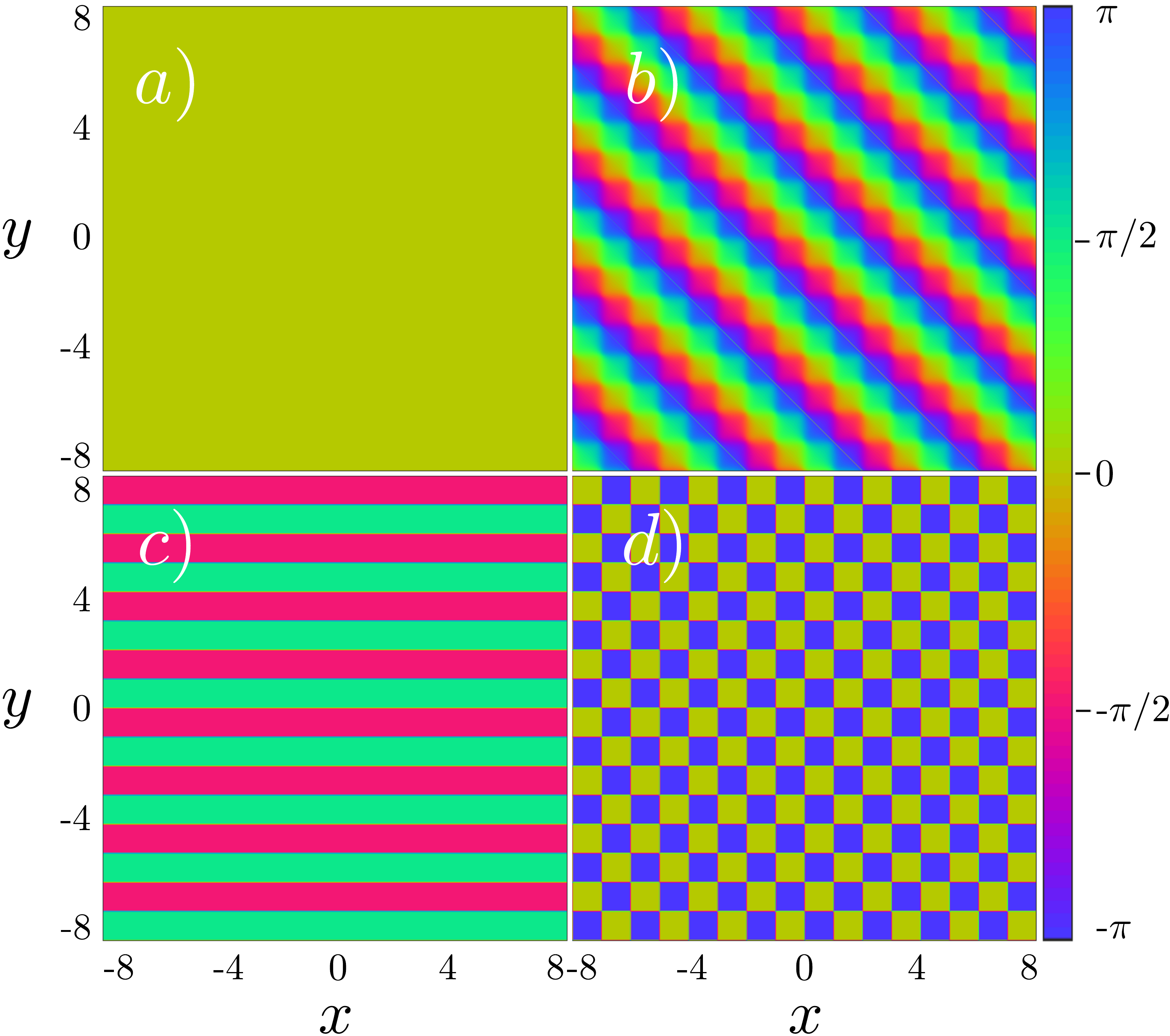}
\end{center}
\caption{Representation of the phase of $\varphi_{\bf Q}(x,y)$ for $V_0=100$ and
$N=16$. 
 The four panels show the same four values of
${\bf Q}$ considered for Fig. \ref{fig1}.
}
\label{fig2}
\end{figure}

\begin{figure}[h!]
\begin{center}
\includegraphics[width=\columnwidth]{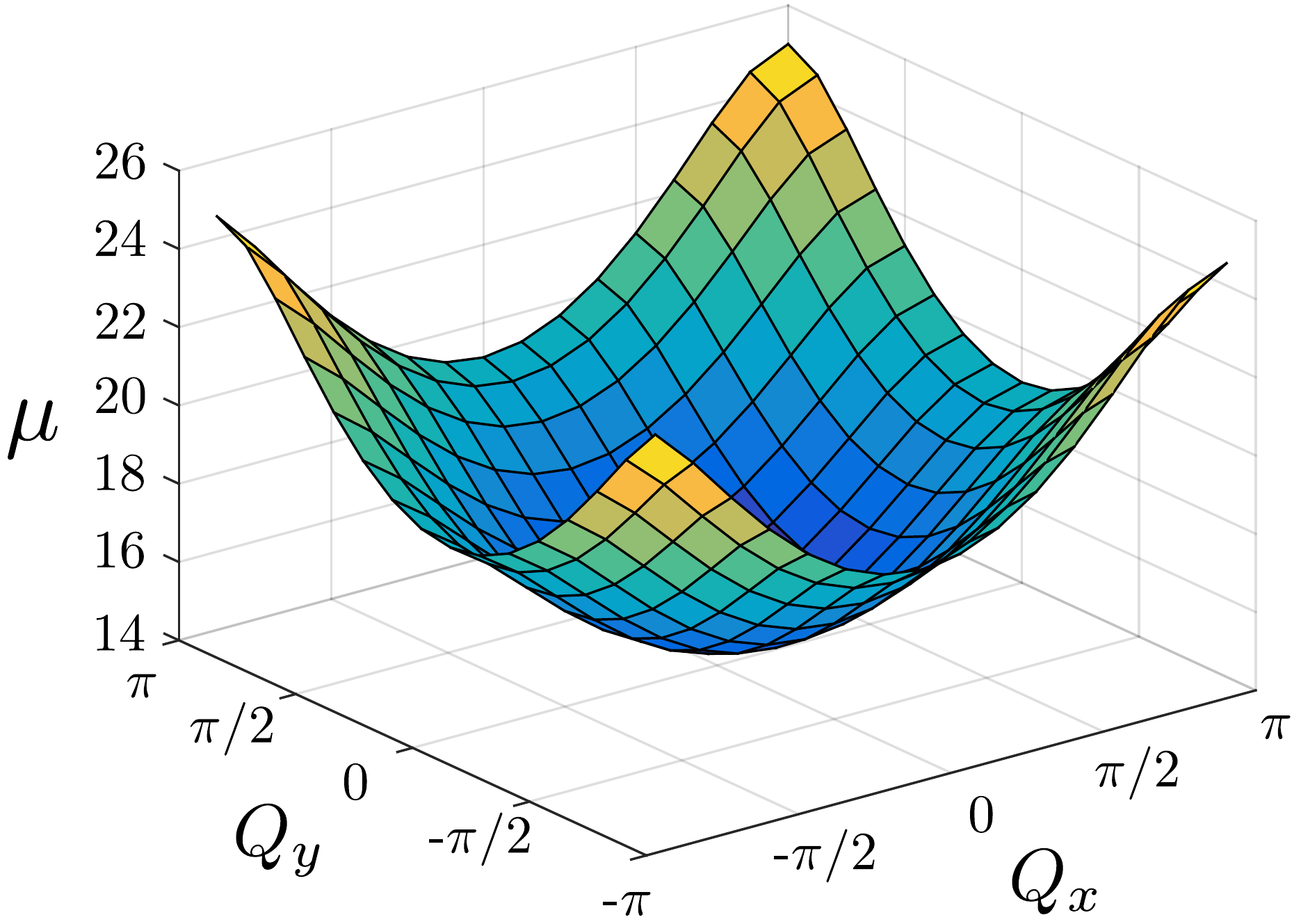}
\end{center}
\caption{
Dispersion relation $\mu(Q_x,Q_y)$ of the lowest band ($V_0=100$).
}
\label{fig3}
\end{figure}

We detail here the linear problem ($P \to 0$ or $g=0$). 
Bloch theorem states that the eigenfunctions can be written in terms of Bloch waves 
$e^{-i\mu_{\bf Q}z}\varphi_{{\bf Q}}({\bf x})$ with
\begin{equation}
\varphi_{{\bf Q}}({\bf x}) = \frac{\sqrt{P}}{N}e^{i {\bf Q}\cdot {\bf x}} u_{{\bf Q}}({\bf x}).
\label{blochlin}
\end{equation}
The $u_{{\bf Q}}$ is a complex function with period 1, namely
$u_{{\bf Q}}(x,y)=u_{{\bf Q}}(x+1,y)=u_{{\bf Q}}(x,y+1)$, which can be computed for 
a single cell by solving
\begin{equation}
\mu_Q u_{\bf Q} = - \nabla^2 u_{\bf Q} + {\bf Q}^2 u_{\bf Q} + V({\bf x}) u_{\bf Q} -2 i {\bf Q} \cdot
\nabla u_{\bf Q},
\label{blochlin2}
\end{equation}
with periodic conditions for $u_{\bf Q}$ at the edges of the unit side cell. 
There should a band index (the generalization of the quantum number for
a single well) but we omit it and restrict ourselves to the
lowest band, an approximation that has been kept throughout the paper. 
Normalization conditions are
\begin{eqnarray}
&&\int_{0}^1 dy \int_0^1 dx |u_{{\bf Q}}({\bf x})|^2 =1, \\
&&\int_\Omega d^2{\bf x} 
\varphi_{{\bf Q_1}}^*({\bf x}) 
\varphi_{{\bf Q_2}}({\bf x}) = P \,\delta_{{\bf Q_1,Q_2}},
\end{eqnarray}
where we have introduced $\Omega$ to refer to the whole domain, namely
$\int_\Omega d^2{\bf x} \equiv \int_{-\frac{N}{2}}^\frac{N}{2} dx  \int_{-\frac{N}{2}}^\frac{N}{2}dy$.
The pseudomomentum ${\bf Q}$ is discretized because of the finite area
and is confined to the first Brillouin cell and, therefore, it can take $N^2$ values
\begin{equation}
Q_x,Q_y \in \frac{2 \pi}{N}\left( -\frac{N}{2} + 1, -\frac{N}{2}+2 ,\dots , \frac{N}{2}-1 , \frac{N}{2}\right),
\label{eq:Q}
\end{equation}
where we have assumed that $N$ is even.
The $N^2$ Bloch functions form a basis for the states in the lowest band. 
They can be readily obtained by standard numerical techniques
from Eq. (\ref{blochlin2}). 
In Figs. \ref{fig1} and \ref{fig2}, we plot some examples of Bloch waves.
In Fig \ref{fig3} we plot
the dispersion relation in the lowest band $\mu({\bf Q})$.

Notice that when  $Q_x$ and $Q_y $ take values 0 or $\pi$, the Bloch functions $\varphi_{\bf Q}(x,y)$ are real.
In particular, $\varphi_{(0,0)}$ is everywhere positive (see panel (a) of Fig. \ref{fig2}) and
 $\varphi_{(\pi,\pi)}$ changes from positive to negative in alternating cells following a chess board
 pattern (see panel (d) of Fig. \ref{fig2}). For $Q_x=0$, $Q_y=\pi$ or viceversa, the phase pattern is striped
 (panel (c) of Fig. \ref{fig2}).
 For the rest of ${\bf Q}$, the Bloch functions
are complex (one could change to a real basis by combining 
$\varphi_{\bf Q}$ and $\varphi_{- {\bf Q}}$).

\begin{figure}[h!]
\begin{center}
\includegraphics[width=\columnwidth]{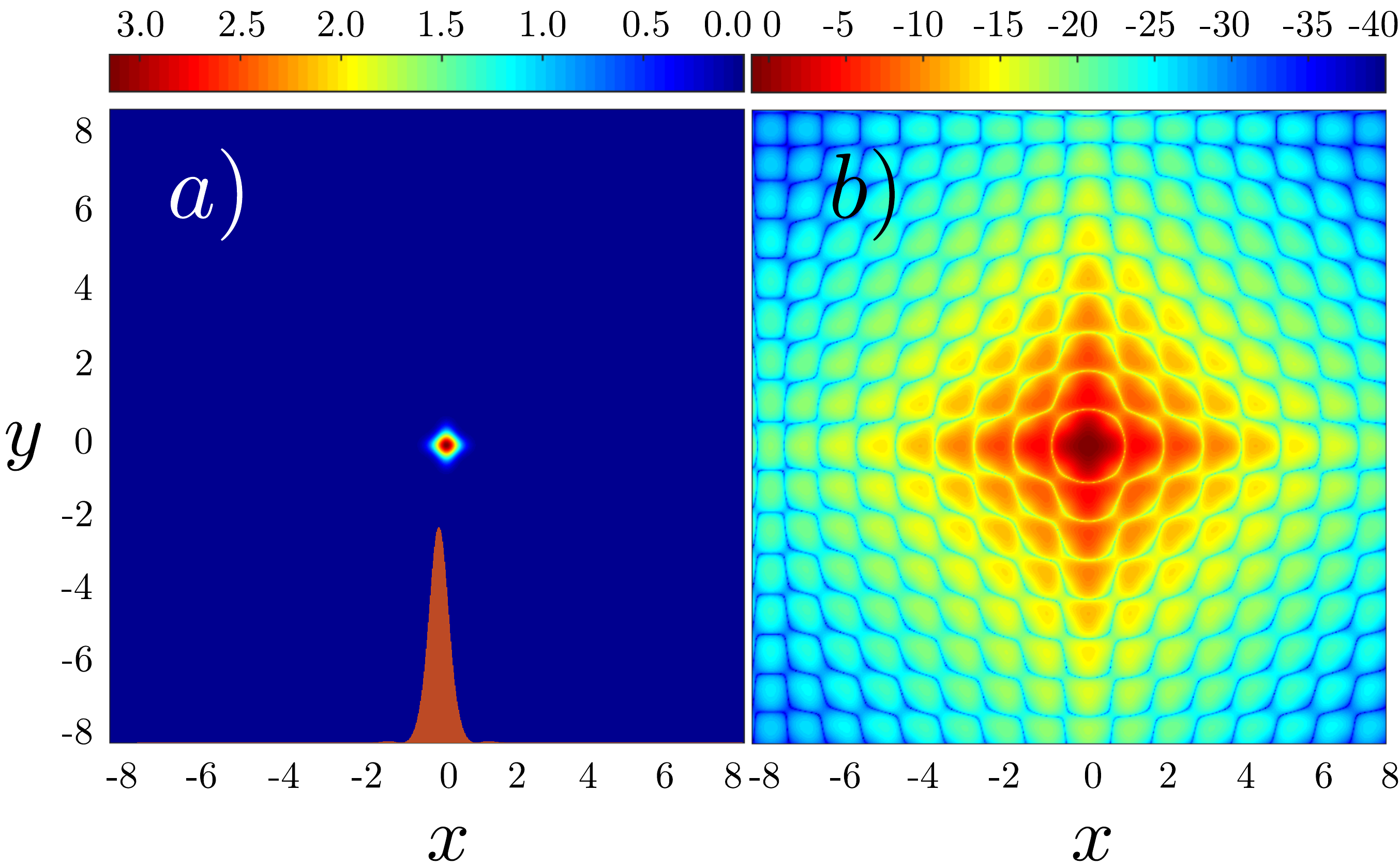}
\end{center}
\caption{Square modulus of the 
Wannier function for ${\bf R}=(-\frac12,-\frac12)$, with $V_0=100$, $N=16$.
Panel (a) is a colormap for $|W_{\bf R}({\bf x})|^2$. 
We also include (in the lower part, in orange) the representation 
of a one-dimensional section $|W_{\bf R}(x,-\frac12)|^2$
 (arbitrary units).
Panel (b) uses a logarithmic scale,
$\ln |W_{\bf R}|^2$. It shows that $W_{\bf R}({\bf x})$ has support in the whole domain but
becomes negligible far from the cell parameterized by ${\bf R}$.
}
\label{fig4}
\end{figure}

Let us now turn to 
the Wannier functions for the lowest band. They constitute a basis of localized functions and
are linear combinations of the 
 Bloch functions,
\begin{equation}
W_{\bf R}({\bf x})=\frac{1}{N\sqrt{P}} \sum_{\bf Q} e^{-i {\bf Q}\cdot {\bf R}} 
\varphi_{\bf Q}({\bf x}),
\end{equation}
where the sum runs over the $N^2$ Bloch momenta and ${\bf R}$  takes values corresponding
to the center of the $N^2$ lattice sites
\begin{equation}
R_x, R_y\in\left(\frac{-N+1}{2}, \frac{N+3}{2}, \dots ,\frac{N-3}{2}, \frac{N-1}{2}\right).
\label{eq:posindexW}
\end{equation}
The prefactor is fixed to normalize the $W_{\bf R}({\bf x})$ as
\begin{equation}
\int_\Omega d^2{\bf x} W_{{\bf R_1}}^*({\bf x})
 W_{{\bf R_2}}({\bf x}) = \delta_{{\bf R_1,R_2}}.
\end{equation}
Given the Wannier function for one of the sites, 
all the rest can be found by a spatial translation $W_{\bf R + n}({\bf x + n})= W_{\bf R }({\bf x })$,
where ${\bf n}$ is a lattice vector.
In Fig. \ref{fig4}, we represent one of them.

The inverse transformation is
\begin{equation}
\varphi_{\bf Q}({\bf x}) = \frac{\sqrt{P}}{N} \sum_{\bf R} e^{i {\bf R} \cdot {\bf Q}} W_{\bf R}({\bf x}).
\label{BlochWann}
\end{equation}

\section{Definition of the phase perturbation at each lattice site}
\label{appendix}

In sections \ref{sec:goldstone} and \ref{sec:finite} we have discussed the propagation of phase
perturbations. It is important to introduce an appropriate prescription to represent it.
For the unperturbed solution, the phase evolves 
 with propagation constant $\mu$ and, 
moreover,  it also changes from cell to cell at any $z$. Directly
plotting the phase is not illustrative. We define here a procedure that is useful to visualize
the dynamics (Figs. \ref{fig8}, \ref{fig10}, \ref{fig11}, \ref{fig12}) and  to compute the velocity.
Other prescriptions are possible, the discussion does not depend on this definition.

First, we assign a phase to each cell by averaging
\begin{equation}
\bar \phi_{\bf R}(z) = \frac{\int d^2{\bf x}  |\psi|^2 {\textrm {arg}}(\psi)}{\int d^2{\bf x}   |\psi|^2},
\end{equation}
 where the integrals are taken within the lattice site with center at ${\bf R}$. 
This average does not make sense in general because of the cyclic 
character of the phase, but is well-defined for all the cases at hand, since the in-site phase variations
are small, cf. Fig. \ref{fig6}. 
Then, we subtract the initial phase of the unperturbed solution and the phase of
a reference cell far from the perturbation.
\begin{equation}
\phi_{\bf R}(z) = \bar\phi_{\bf R}(z) - \bar\phi_{\bf R}(0) - (\bar\phi_{{\bf R}_{ref}}(z) -
 \bar\phi_{{\bf R}_{ref}}(0) ). 
\end{equation}

With this prescription, the $\phi_{\bf R}(z)$ of a stable unperturbed solution is zero for all $z$.
For a perturbed solution, it remains zero far from the perturbation, at least until it reaches
${\bf R}_{ref}$.

\acknowledgments

MAGM and AF acknowledge fruitful discussions with Thomas Busch and Yongping Zhang. 
This work is supported by grants FIS2014-58117-P and TEC2014-53727-C2-1-R
from Ministerio de Econom\'\i a y Competitividad (Spain) and grant GPC2015/019
from Xunta de Galicia. MAGM acknowledges  support from ERC Advanced Grant OSYRIS, EU IP SIQS, EU PRO QUIC, EU STREP EQuaM (FP7/2007-2013, No. 323714), Fundació Cellex, the Spanish MINECO (SEVERO OCHOA GRANT SEV-2015-0522, FISICATEAMO FIS2016-79508-P), and Generalitat de Catalunya (SGR 874 and CERCA/Program).

\end{document}